\documentclass[traditabstract]{aa}

\usepackage{graphicx}
\usepackage{txfonts}
\usepackage{color}
\usepackage{natbib}
\usepackage{epsfig}
\usepackage{psfrag}
\usepackage{lineno}
\usepackage[toc,page]{appendix}
\usepackage{hyperref}
\hypersetup{colorlinks   = true,
	    citecolor    = blue, 
	    linkcolor	 = blue}

\begin{document}
   \title{Exploring the nature of broadband variability in the FSRQ 3C~273}

   \titlerunning{Broadband variability in  3C~273}

   \author{
          C. Chidiac\inst{1}\thanks{Master student at Argelander Institut f{\"u}r Astronomie (AIfA) - University of Bonn}
          \and B. Rani \inst{1}
          \and T. P. Krichbaum\inst{1}
          \and E. Angelakis\inst{1}
          \and L. Fuhrmann\inst{1}
          \and I. Nestoras\inst{1}
          \and J. A. Zensus\inst{1}
          \and A. Sievers\inst{2}
          \and H. Ungerechts\inst{2}
          \and R. Itoh\inst{3}
          \and Y. Fukazawa\inst{3}
          \and M. Uemura\inst{4}
          \and M. Sasada\inst{5}
          \and M. Gurwell\inst{6}
          \and E. Fedorova\inst{7} \\
         }
   \institute{
              Max-Planck-Institut f{\"u}r Radioastronomie (MPIfR), Auf dem H{\"u}gel 69, D-53121 Bonn, Germany
	    \and {Instituto de Radioastronom{\'i}a Milim{\'e}trica  (IRAM), Avenida Divina Pastora 7, Local 20, 18012 Granada, Spain}
	    \and { Department of Physical Science, Hiroshima University,
		  Higashi-Hiroshima, Hiroshima 739-8526, Japan}
	    \and {Hiroshima Astrophysical Science Center, Hiroshima University,
	    Higashi-Hiroshima, Hiroshima 739-8526, Japan}
	    \and Institute for Astrophysical Research, Boston University,
	      725 Commonwealth Avenue, Boston, MA 02215, USA
	    \and Harvard-Smithsonian Center for Astrophysics, Cambridge, MA 02138, USA
	    \and Astronomical Observatory of National Taras Shevchenko University of Kiev}
	    
	    \authorrunning{Chidiac et al. 2016}

   \date{Received ---------; accepted ----------}

\abstract
{Detailed investigation of broadband flux variability in the blazar
3C~273 allows us to probe the location and size of emission regions and
their physical conditions. We report the
results on correlation studies of the flaring activity in 3C~273 observed
for a period between 2008 and 2012. The observed broadband
variations were investigated using the structure function and the
discrete correlation function methods. Starting
from the common use of power spectral density  analysis (PSD) at X-ray frequencies,
we extended our investigation to characterize the nature of variability at radio, optical, 
and $\gamma$-ray frequencies. The PSD analysis showed that the optical/IR light curve slopes are consistent
with the slope of white noise processes; while, the PSD slopes at radio,
X-ray and $\gamma$-ray energies are consistent with red-noise processes. We found that 
the estimated fractional variability amplitudes have a strong dependence on the observed frequency. 
The flux variations at $\gamma$-ray and mm-radio bands are found to be significantly correlated. Using the 
estimated time lag of (110$\pm$27) days between $\gamma$-ray and radio light curves  where $\gamma$-ray 
variations lead the radio bands, we constrained the
location of the $\gamma$-ray emission region at a de-projected
distance of $1.2\pm0.9$~pc from the jet apex. Flux variations
at X-ray bands were found to have a significant correlation with variations
at both radio and $\gamma$-rays energies. The correlation between
X-rays and $\gamma$-rays light curves provides a hint of two possible
time lags, which suggests presence of two components responsible for the X-ray emission. A
negative time lag of  -(50$\pm$20)~days,  where the X-rays are leading the emission, 
suggests X-rays are emitted closer to the jet apex from a compact
region (0.02--0.05~pc in size) i.e.\ most likely from the corona at a distance
of (0.5$\pm$0.4)~pc from the jet apex. A positive time lag of
(110$\pm$20) days ($\gamma$-rays are leading the emission) suggests jet-base origin of the 
other X-ray component at $\sim$ 4--5~pc from the jet apex. The flux variations at radio 
frequencies were found to be well correlated with each other such that the variations at higher 
frequencies are leading the lower frequencies,  which could be expected in the standard 
shock-in-jet model.}


\maketitle

\section{Introduction}

The Flat Spectrum Radio Quasar (FSRQ) 3C~273 is one
of the most studied  quasars since its discovery in 1963. It is a nearby
source, with a red shift of 0.158 (\citealt{Schmidt1963}).
It is characterised by a jet close to the line-of-sight  at
a viewing angle of $5^{\circ} - 11^{\circ}$ (\citealt{Liu2009})
and a maximum superluminal apparent velocity of $v_{\mathrm{app}} = 15$ c
(\citealt{Lister2013}). The jet, though  extending
up to kpc scales,  remains collimated  in
all wavebands, from radio to X-rays,
allowing  the detailed study of the physical properties of the AGN
starting from  regions close to the supermassive black hole (SMBH)  to
the outer jet regions (\citealt{Uchiyama2006}).

The blazar 3C~273 is characterised by a strong radio
emission from mm- to cm-wavelengths (\citealt{Soldi2008}). It has
been suggested that this radio emission is caused 
by synchrotron emission of relativistic electrons in the jet,
that could emit up to infra-red (IR) frequencies (\citealt{McHardy1999,Soldi2008}).
\citealt{Turler1999}  found that the sub-mm/radio flaring activity of
the source is in agreement with the shock-in-jet model suggested by \citealt{Marscher1985}. 
They also found that the high radio frequency outbursts are short-lived and are
coming from a region closer to the
jet apex than the long-lived low radio frequency outbursts.

Unlike other blazars that are highly polarised,  3C~273 has an average polarisation
degree below 1\% at optical frequencies (\citealt{Valtaoja1991}).
The source is characterised by a blue bump due to the excess
in the optical/UV emission. Two possible components have been suggested to explain the complex 
optical/UV spectrum. One rapidly variable component could be from the accretion
disk (\citealt{Shields1978,Soldi2008}), or illumination by an X-ray
source (\citealt{Ross1993}) or a hot corona (\citealt{Haardt1994}).
The second component seems to be related to synchrotron emission from 
the jet (\citealt{Paltani1998}). 
The IR emission in 3C~273 seems   to be due to two variable components, (\citealt{Paltani1998}),
where one component is due 
to thermal emission from the dusty torus and the other
component is due to synchrotron emission from the jet (\citealt{Robson1993,McHardy1999,Sokolov2005}).

The source exhibits significant variations in  the
X-ray regime. Several studies have been carried out in order to
understand the emission mechanisms and locate the radiation regions. Studies by \citealt{McHardy1999,Grandi2004,McHardy2007}
suggested that a Seyfert-like component from the accretion disk is
responsible for around 20\% of the X-ray emission, while the rest could
be due to a synchrotron self-Compton process, most likely, the up-scattering
of the IR photons that had been produced by the synchrotron emission
of the jet. Alternatively,  the X-ray emission could be due to the
comptonisation of the UV photons (\citealt{Madsen2015}). 
 
The presence of a weak and neutral iron line in
the X-ray spectrum suggests that the X-ray emission arises from the
corona and reflection off the accretion disk (\citealt{Madsen2015}). The source therefore exhibits Seyfert-like properties in the X-ray regime
(\citealt{McHardy1999,Grandi2004}).

A previous study carried out on the multi-wavelength variability of
3C~273, found a correlation between the X-ray and
the IR flares with the latter leading the former by a time lag
of ($0.75\pm 0.25$) days (\citealt{McHardy1999}),
which rules out the External Compton (EC) emission process as possible
producer of X-Ray emission. This correlation was later confirmed by
\citealt{McHardy2007}. \citealt{Chernyakova2007} found a correlation
between UV and X-rays; however, this correlation is still debatable
since there are studies that  failed to confirm it (\citealt{Walter1992,Soldi2008}).
Many studies found  for some sources a strong correlation between radio and $\gamma$-rays
(\citealt{Beaklini2014,MaxMoerbeck2014,Ramakrishnan2015}), suggesting
thus a common mechanism that is responsible for both radio and $\gamma$-rays.  Moreover, the gamma-ray flares were accompanied by ejection 
of new components from the base 
of the jet \citep{Jorstad2012}. 

The source was first detected in $\gamma$-ray frequencies by COS-B
in 1970.   Later,  Energetic Gamma-ray Experiment Telescope (EGRET) on board 
Compton Gamma-Ray Observatory (CGRO) detected the source in 1991 at energies higher
than 100 MeV \citep{Hartman1992}. {\it Fermi}/LAT (Large Area Telescope) detected 3C~273 since the beginning
of its operation in 2008 (\citealt{Abdo2010b}). In the period between
July 2009 and April 2010, a strong flaring activity  on
GeV scales was reported. A ten-day flare was observed in August 2009
followed by two bright flares on 15-19 and 20-23 September 2009, which was followed by a sequence of rapid flares. The
last flare was observed in April 2010 and since then the source went
into a quiescent state that is lasting until the present day (\citealt{Rani2013}).
The fastest $\gamma$-ray flare had a doubling time-scale of 1.1 hour
(\citealt{Rani2013}). 
Following the episodes of extreme flaring activity
in the source since 2008 \citep{Rani2013}, we investigate the nature 
of the observed broadband flaring activity. In
this paper, we carried out a detailed cross-correlation investigation
across the electromagnetic spectrum 
to provide better constrains on the emission mechanisms. The key objective is to provide a better constrain on the location and 
size of emission region.
This paper is structured as follows. In section $\ref{data}$,
we explain the data used in this study. Section
$\ref{analysis}$ presents the light curves, and we explain the methods
applied in this paper, and the results of each method follow
respectively. A discussion of the results is given in section $\ref{discussion}$. A conclusion is given in section $\ref{conclusion}$.

\section{Observations and data reduction}
$\label{data}$
To explore the broadband flaring activity, we monitored the source
using both space- and ground-based observing facilities for a time period
between  May 2008 and  March 
2012 (MJD= 54600 to 56000). In the following subsections, we summarise the observations and data reduction.

\subsection{Gamma-ray data}
The high energy GeV observations used in this study were obtained
in a survey mode by {\it Fermi}-LAT. 
The data covered an energy range from $100\, \mathrm{MeV}$
to $300\, \mathrm{GeV}$. Here we use the weekly and monthly
 averaged data of 3C~273 that are already
presented in \citealt{Rani2013} where the details of the observations and data reduction
 are discussed.

\subsection{X-ray data}

X-ray observations of the source were obtained by two space-based
instruments:  SWIFT and Rossi X-ray Timing Explorer (RXTE). The X-ray light
curves cover three energy bands. The monthly averaged hard X-ray light curve,
covering  an energy range between $14-192\,\mathrm{keV}$  was observed
by Swift-Burst Alert 
Telescope (BAT)\footnote{http://swift.gsfc.nasa.gov/results/bs70mon/$\newline$SWIFT\_J1229.1p0202} 
in both photon counting (PC) and windowed timing (WT) modes. Details
of the observations and data reduction\footnote{http://www.swift.psu.edu/monitoring/readme.php} 
are given in \citealt{Stroh2013}. Observations at $2-10\, \mathrm{keV}$  and $10-50\, \mathrm{keV}$ 
X-ray bands are provided by Rossi X-ray Timing Explorer - Proportional
Counter Array (RXTE-PCA)\footnote{http://heasarc.gsfc.nasa.gov/docs/xte/PCA.html}.  
The data are available for public use\footnote{http://www.swift.psu.edu/monitoring/source.php?$\newline$source=3C273$\newline$http://cass.ucsd.edu/\textasciitilde{}rxteagn/3C273/3C273.html}. Details about the instrument and data 
reduction\footnote{http://cass.ucsd.edu/\textasciitilde{}rxteagn/} are presented in \citealt{Jahoda1994}.

\begin{figure*}[!t]
\center
\includegraphics[scale=0.7]{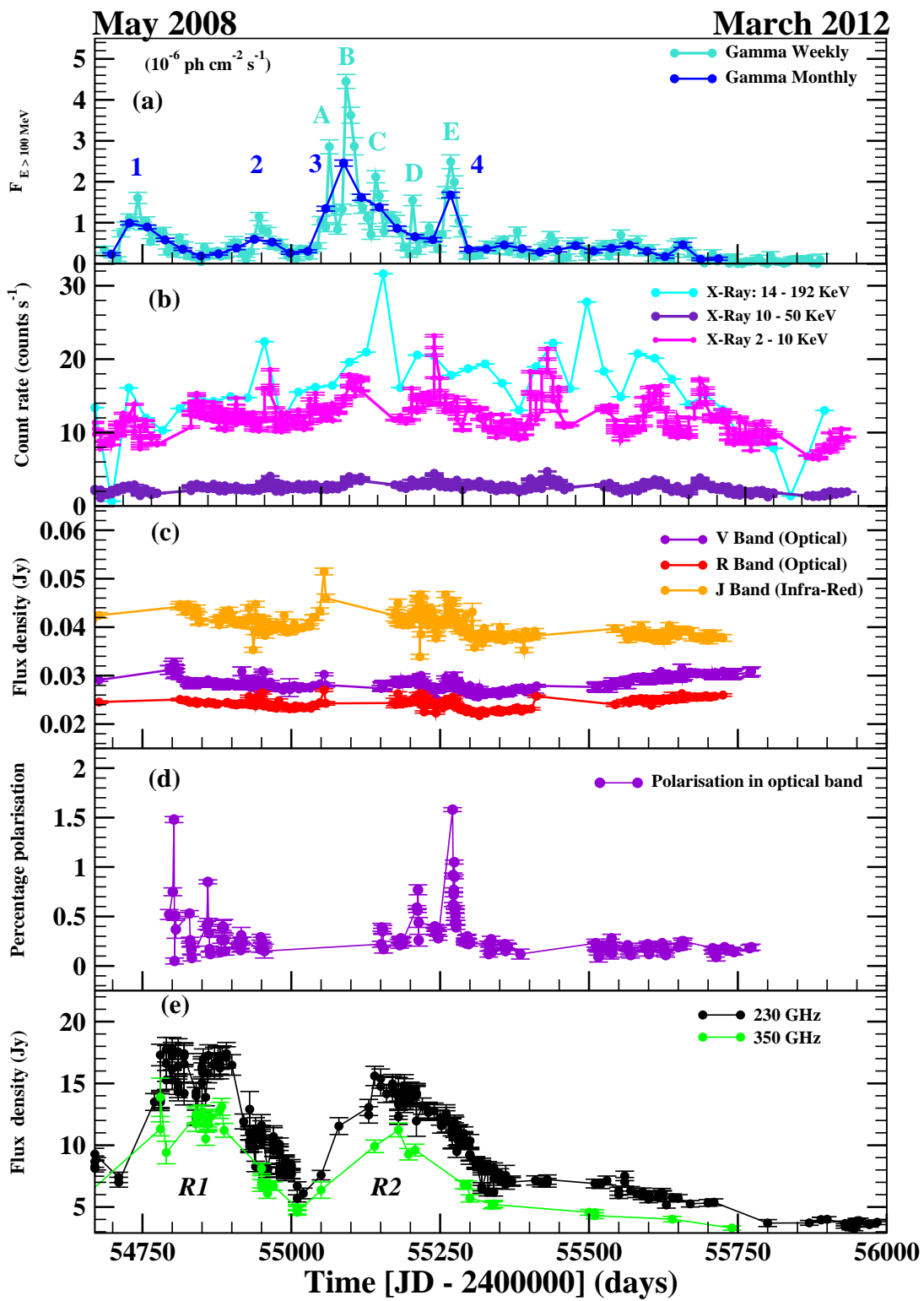}
\caption{Broadband light curves of 3C~273: (a) weekly (turquoise
circles) and monthly (blue circles) averaged $\gamma$-ray light curves; (b) 14-192 keV band (cyan), the 10-50 keV band (indigo)
and the 2-10 keV band (magenta) X-rays light curves; (c) optical (V and R passbands) and IR (J passband) light curves; (d)   optical percentage polarisation 
curve; and (e) radio flux density
curves at 230~GHz (in black) and 350~GHz (in green) bands.}
\label{LC_all}
\end{figure*}

\subsection{Optical and IR data}
Optical and IR light curves were provided by  Small and
Moderate Aperture Research Telescope System (SMARTS) monitoring 
programme\footnote{http://www.astro.yale.edu/smarts/glast/home.php}. 
SMARTS  observes with telescopes located
at Cerro Tololo Inter-american Observatory (CTIO). The monitoring programme aims 
to understand the high energy emission of blazars by searching for temporal
correlation between flux and spectra  of
the major emission components. The
observation and data reduction of the light curves can be found in
\citealt{Bonning2012}. Observations by the 2.3\,m Bok Telescope of Steward 
Observatory\footnote{http://james.as.arizona.edu/\textasciitilde{}psmith/{\it Fermi}/} 
complemented the optical data. The Steward observatory provides optical
data in R and V bands for blazars and {\it Fermi} targets of opportunity
sources. Details about the light curves produced in Steward Observatory
can be found in \citealt{Smith2009}.  The polarisation measurements are derived from the median
Stokes Q and U values found from spectropolarimetry in a 5000-7000\,\AA{} bin.
We also used optical photometric data from the Kanata Telescope
in Hiroshima Observatory\footnote{http://hasc.hiroshima-u.ac.jp/telescope/kanatatel-e.html}. 
Details about the telescope and its instruments
are referred to \citealt{Uemura2009}. Details
of the observation and data reduction of the light curves can be
found in \citealt{Ikejiri2011}.

\subsection{Radio data}
Radio band observations at 2.6, 5, 8,
10, 15, 23, 32, 43, 86, and 142~GHz were provided by the
FERMI-GST AGN Multi-frequency Monitoring Alliance (F-GAMMA) programme%
\footnote{http://www3.mpifr-bonn.mpg.de/div/vlbi/$\newline$fgamma/fgamma.html%
} (\citealt{Fuhrmann2007,Angelakis2008}). The F-GAMMA programme 
uses  the Effelsberg 100\,m telescope that
covers a range from 2.6 to 43~GHz and   the IRAM 30\,m telescope at 86 and 142\,GHz at Pico
Veleta Observatory. The
millimetre observations are closely coordinated with the more general
flux density monitoring conducted by IRAM, and data from both programmes
are included in this paper. The 230
and 350~GHz light curves are given by the SMA Observer Centre
\footnote{http://sma1.sma.hawaii.edu/callist/callist.html
} data base (\citealt{Gurwell2007}).

\section{Analysis and results }
$\label{analysis}$
In this section, we present the statistical analysis of the observed
broadband variations in 3C~273. We compare the fractional
variability in each band as well as the variability time scale. To quantify
the correlation  between different frequencies, we   employed the cross-correlation
function method. Finally, we used the Power Spectral Density (PSD) method
to get a better insight on the nature of the variability by describing
the amount of variability as function of temporal frequency. The significance
of the obtained results was tested via simulations. In the following sub-sections,
we discuss the analysis in details.

\subsection{Light curves analysis}
\begin{figure}
\centering{}\includegraphics[scale=0.35, trim=1 1 1 1.3, clip=true]{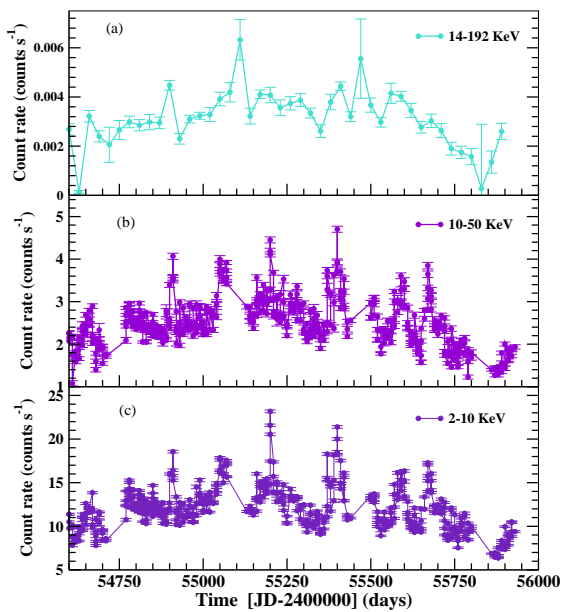}
\caption{X-ray light curves of 3C~273 at  14$-$192 $\mathrm{keV}$ (a), 10$-$50 $\mathrm{keV}$ (b), and 
 2$-$10 $\mathrm{keV}$ (c). }
\label{xrayLC}
\end{figure}

\begin{figure}
\begin{centering}
\includegraphics[width=8cm,height=10cm,keepaspectratio]{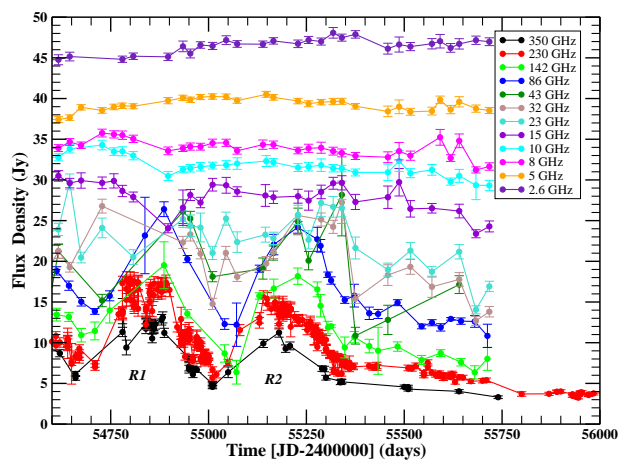}
\par\end{centering}
\caption{$\label{radioLC}$ Radio (cm to mm) bands light curves of 3C~273.
Different colours represent different frequencies, from 2.6~GHz to
350~GHz (from top to bottom). }
\end{figure}

Figures $\ref{LC_all}, \ref{xrayLC}$ and \ref{radioLC} show the broadband light curves of 3C~273.
Panel (a) in Fig. \ref{LC_all}, shows the monthly averaged $\gamma$-ray light curve  superimposed on top 
of the weekly averaged curve. 
The monthly sampled light curve shows fewer flares (labelled as 1 to 4) compared to the weekly
sampled light curve. Apparently, the source seems to exhibit two different modes of flaring 
activity. A mode of slow activity (between MJD =
54600 to 55000) is followed by
a second phase that is characterised by rapid and strong flaring activity (flare A to E) between 
55050 and 55300 MJD. 
The source went to a quiescent state after flare E observed in April 
2010, which is still  persisting. 
The source also
exhibited significant spectral variations during the flaring activity.
At the beginning of each flare, the spectrum  becomes harder, then softens
again, at the end of the flare \citealt{Rani2013}.

Similarly to the $\gamma$-ray regime, the source was active in X-ray
bands (see Fig.$\ref{LC_all}$ b). Apparently,   the
flaring activity seems to be very similar in the three observed X-ray
bands (Fig.\ref{xrayLC}). 
As seen for the
Swift-BAT light curve, there is a major flare that was observed almost
simultaneously with flare (C) in the $\gamma$-ray flares. 
However, when the source went to a quiet state at $\gamma$-ray regime, it continued 
flaring at X-ray frequencies.

Figure $\ref{LC_all}$ (c) shows the optical (R and V passbands) and
IR (J passband) light curves. Unlike  at
$\gamma$-ray and X-ray energies, the optical/IR light curves of the source
show lower amplitude fluctuations without
pronounced peaks. The light curves have an almost constant flux during the entire
period of observation.  
In addition to that the observed optical  fractional 
polarisation is quite low (<2\%). We however noticed significant variations in the polarisation 
curve (Fig.\ 1 d),  indicative
of an underlying non-thermal variable emission component. Apparently,
the variations in optical polarisation became more pronounced during
the $\gamma$-ray flaring activity. When the source went into a quiescent
state  in the GeV regime, no pronounced activity is noticed in the percentage
polarisation curve. 

Figure $\ref{LC_all}$ (e)  shows the light curves at 230 and 350~GHz, 
for the  comparison of the variability of the mm-band
with the variability in other energy bands. Figure $\ref{radioLC}$
shows all the radio band light curves of the source. Higher radio
frequencies from 23~GHz to 350~GHz show two flares: flare $\mathcal{R}_{1}$
between 54600 and 55050 MJD, and flare $\mathcal{R}_{2}$ between
55050 and 55600 MJD. At 230 and 350\,GHz radio bands, we noticed sub-flaring activity on short time scales, 
superimposed on top of the major flare
$\mathcal{R}_{1}$. This flare is observed simultaneously with the flares 1 and 2 in GeV regime while
flare $\mathcal{R}_{2}$ is observed during the strong flaring period
between 55000 and 55300 MJD at $\gamma$-rays frequencies. The flaring
activity is less pronounced at frequencies  below 23~GHz, which
could be due to opacity. It is important to note that when the source was quiet in  the
GeV regime, the radio light curves did not show any prominent
variations.

\subsection{Fractional variability}
\begin{figure}
\begin{centering}
\includegraphics[scale=0.34]{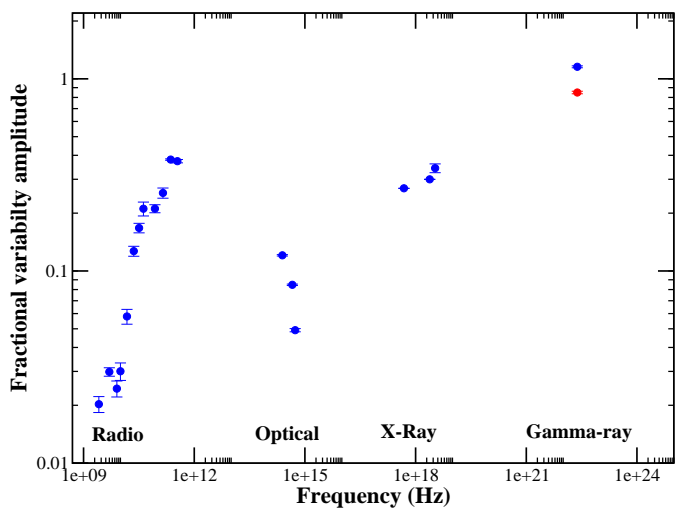}
\end{centering}
\caption{Fractional variability amplitude $\label{fracplot}$ plotted as a
function of frequency. The red point describes the fractional variability amplitude of the monthly averaged $\gamma$-ray light curve.}
\end{figure}

We compute the fractional variability amplitude to compare the variations at different energy bands.  Following
\citealt{Vaughan2003}, we define the fractional variability as follows:
\begin{equation}
F_{var}=\sqrt{\frac{S^2-\overline{\sigma_{err}^2}} {\bar{x}^{2}}} \label{fracvar}
\end{equation}
where $S^2$ is the variance of the light curve, $\bar{x}$  is the
mean value of the flux in the light curve, and $\overline{\sigma^2_{err}}$ 
its mean square error. The formal error
of the fractional variability is estimated by (\citealt{Vaughan2003},
Appendix B):
\begin{equation}
err(F_{var})=\sqrt{\frac{1}{2N}\left(\frac{\overline{err^{2}}}{F_{var}\times\overline{x}^{2}}\right)^{2}+\frac{\overline{err^{2}}}{N}\times\frac{1}{\overline{x}^{2}}}
\end{equation}
where $\overline{err^{2}}$ is the error on the flux. We adopted this method because it takes the uncertainty on the flux
in account for calculating the variability amplitude as well as the error on it. 
By applying Eq. (\ref{fracvar}), we obtained an estimate of the
fractional variability of all the light curves available for 3C~273. 
The estimated $F_{var}$ amplitudes for the broadband light curves are plotted 
in Fig. $\ref{fracplot}$. 

As seen in Fig. $\ref{fracplot}$, the
variability amplitude for the radio light curves tends to increase
with increasing frequency. This could
be explained in terms of opacity effects i.e.\ at higher frequencies, we are probing 
the optically thin regions in the  jet. 
There is  less significant variability in the optical and
IR light curves; as a consequence, the estimated fractional variability amplitudes
at these bands are comparatively low. This could 
also be caused by the lack of data around the GeV flaring times.
The fractional variability amplitude increases again in the X-ray regime,
and is comparable to the fractional variability amplitude
at mm-radio bands. The estimated $F_{var}$ amplitudes are significantly higher for the 
$\gamma$-ray light curves exceeding 100\%.

\subsection{Structure function}

\begin{figure}
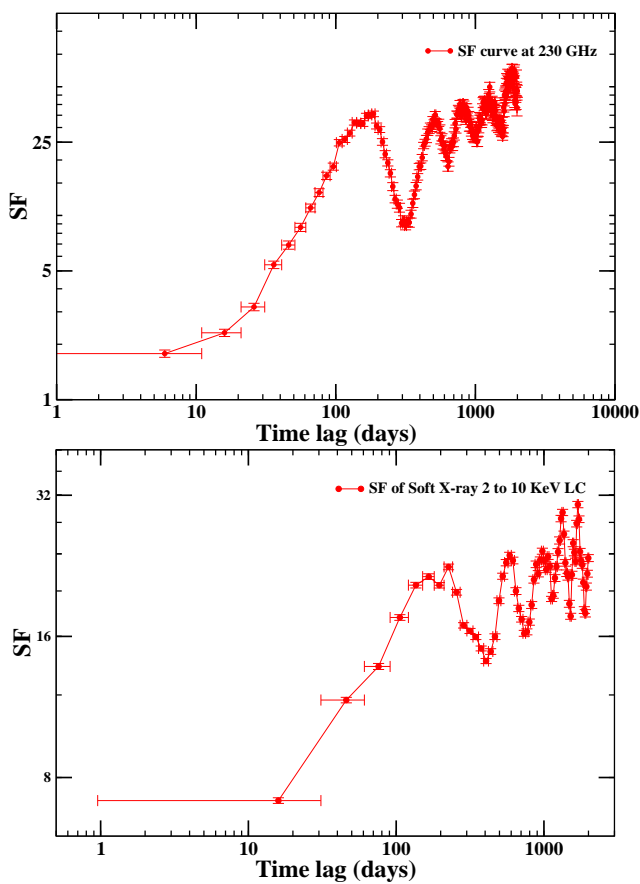

\includegraphics[scale=0.34, trim=1 1 1 1.5, clip=true]{SF_230}
\includegraphics[scale=0.34, trim=-3.5 1 1 1.5, clip=true]{SF_SoftXray2to10KeV}
\caption{Structure function at 230~GHz radio band (top) and 2-10 keV X-ray band (bottom).}
$\label{SF_curve}$
\end{figure}

We used the structure function  (SF) method \citep{Simonetti1985} to calculate the
characteristic variability time scales in the light curves. For a given time series, the structure function \citep{Simonetti1985} is defined as: 
\begin{equation}
SF(\Delta t)=\frac{1}{N}\sum_{i=1}^{N}\left[x_{i}-x_{i+\Delta t}\right]^{2}
\end{equation}
where  $x_{i}$ is the flux at a given time, and
$x_{i+\Delta t}$ is the flux at a time separation of $\Delta t$, and 
$N$ is the number of data points that share a same $\Delta t$ separation  within a binning time interval. It is important to note that the SF breaks are not always 
reliable indicators of variability timescales \citealt{Emmanoulopoulos2010}. We therefore used the SF method only for comparison
purposes; the estimated timescale are not used for any further calculations.
Using eq.\ 3 we extracted the variability for all the light curves used in this study.
The SF curves at 230~GHz and 2-10 keV bands are shown in Fig.~$\ref{SF_curve}$; for the rest the 
SF curves are shown in Fig.~\ref{fig_app_3}.

As we can see for the 230~GHz SF curve (Fig.\ \ref{SF_curve}), the SF values increase as
a function of time lag reaching a peak which is followed by a dip. The first SF peak corresponds 
to the shortest variability time scale and the dip indicates a possible time scale of periodic 
variation, if it is repeated \citep[see][for details]{Rani2009}. The estimated variability time scales at different 
energy bands are listed in  Table $\ref{SF_table}$.  For
the error on the time lag, we adopted half of the time binning as
a conservative estimate.

\begin{figure}
\begin{centering}
\includegraphics[scale=0.34, trim=1 1 1 1.5, clip=true]{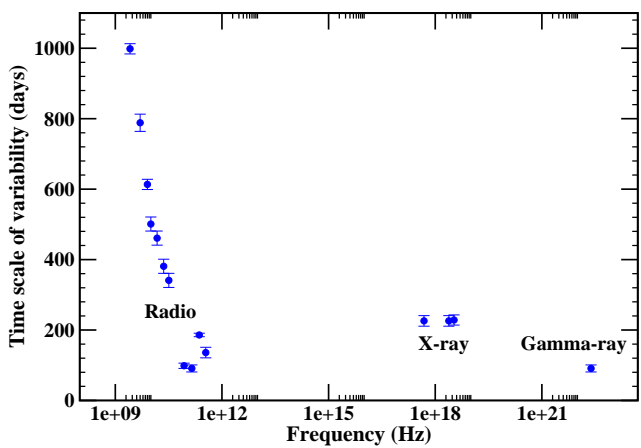}
\end{centering}

\caption{Variability time scales plotted as a function of frequency (Hz).}
$\label{Timescale_variability}$
\end{figure}

As we can see in  Fig.\ \ref{Timescale_variability},  the variability time scales at radio bands decrease 
as we go to higher frequencies. This implies that we are probing compact emission regions 
at short-mm bands, which are opaque to cm radio bands. In the optical bands, the
fractional variability is very low. Therefore,
the variability time scales cannot be determined with accuracy. 

The  variability time scales for the three
X-ray light curves are very similar. For the light curves at 2-10
keV and 10-50 keV bands, we found a time scale of  
(226$\pm$15) days, and for the X-ray bands at 14-192 keV, we found
a time scale of  (228$\pm$15) days. These X-ray variability time scales are comparable to 
the variability time scales at mm-radio bands. In addition to that 
we noticed multiple cycles in the SF curves at both mm-radio and X-ray bands (see Fig. $\ref{SF_curve}$) which suggests 
nearly periodic variations at time scales of 350\,--\,400~days. Comparable time scales for variability and  
quasi-periodic oscillations at short-mm radio and X-ray regimes suggest a possible correlation between 
the two.  

Using the SF analysis of the weekly averaged $\gamma$-ray light curve, we found a variability time 
scale of  (91$\pm$10)\,days. However, it is important to note 
that the SF gives the time scale of the long term behaviour of the light curve.
Interestingly, the estimated $\gamma$-ray variability time scale is 
comparable to that at 86 and 142~GHz radio bands (see Table 1), which could be a hint 
of a possible correlation between the radio and $\gamma$-ray bands. In Section \ref{DCF}, we investigate that in detail.

\begin{table}
\centering 
\caption{Variability time scales as derived from the structure function analysis }
$\label{SF_table}$
\begin{tabular}{cc}
\hline 
 Light curves &  Variability time scale (days) \\\hline 
2.6~GHz & 998$\pm$14\\
5~GHz & 788$\pm$24\\
8~GHz & 613$\pm$14\\
10~GHz & 500$\pm$20\\
15~GHz & 460$\pm$20\\
23~GHz & 380$\pm$20\\
32~GHz & 340$\pm$20\\
86~GHz & 98$\pm$7\\
142~GHz & 90$\pm$10\\
230~GHz & 185$\pm$5\\
350~GHz & 135$\pm$15\\
2-10 keV & 225$\pm$15\\
10-50 keV & 225$\pm$15\\
14-192 keV & 228$\pm$14\\
Gamma-ray & 90$\pm$10\\
\hline 
\end{tabular} 
\end{table}

\subsection{Power Spectrum Density (PSD)}

To explore the nature of the variability, we used the Power Spectrum
Density (PSD) analysis. The PSD method is explained in details in \citealt{Vaughan2003}
and \citealt{Vaughan2005}. For an evenly sampled data set, the periodogram
is the modulus-squared of the Discrete Fourier Transform (DFT) of
the data. In a light curve with flux series $x_i$ measured at discrete
times $t_i$ ($ i = 1, 2,..., N$), the modulus-square of the DFT
is given as: 
\begin{equation}
\left|DFT(f_{j})\right|^{2}=\left|\sum_{i=1}^{N}x_{i}e^{2\pi if_{i}t_{i}}\right|^{2}
\end{equation}
at $ \frac{N}{2}$ evenly spaced temporal frequencies $f_j = \frac{j}{N \Delta T}$
where $ j = 1,2, ... \frac{N}{2}$. The Nyquist frequency is: $f_{Nyq} = \frac{1}{2\Delta T} $.
The periodogram ${I}(f_{j})$ is given by: 
\begin{equation}
{I}(f_{j})=A\left|DFT(f_{j})\right|^{2}=\frac{2\Delta T}{N}\left|DFT(f_{j})\right|^{2}
\end{equation}
The periodogram will be scattered around the true underlying power
spectrum, if we are observing a noise process, following a $\chi^2$
distribution with two degrees of freedom.  The true power spectrum
${P}(f_j)$ is calculated as:
\begin{equation}
I(f_{j}) = {P}(f_{j})\frac{\chi_{2}^{2}}{2}
\end{equation}

\subsubsection{Fitting the periodogram}

The underlying  power density spectrum
can be described  by a power law ${P} (f) = N f^{-a}$
where $a$ is the slope and $ N$ is the normalisation constant.
 We fitted the power law to the PSD using
a least square fit method (LS) in log-scale. The scatter is multiplicative
in linear-space, therefore it is additive in log-space, and identical
at each frequency (\citealt{Geweke1983,Papadakis1993}): 
\begin{equation}
log\left[I(f_{j})\right]=log\left[{P}(f_{j})\right]+log\left[\frac{\chi_{2}^{2}}{2}\right]
\end{equation}
Nonetheless, the expectation value of the periodogram in log-space
is not the same as the expectation value of the spectrum. There is
a bias between the two values. This bias is a constant due to the
shape of the $\chi_2 ^2$-distribution in log-space and it can be
trivially removed. 
\begin{equation}
\left\langle log\left[I(f_{j})\right]\right\rangle =\left\langle log\left[\ensuremath{{P}(f_{j})}\right]\right\rangle +\left\langle log\left[\ensuremath{\frac{\ensuremath{\chi_{2}^{2}}}{2}}\right]\right\rangle 
\end{equation}
From \citealt{Abramowitz1964}, the estimate of this bias constant
is then: $\left\langle log\left[\ensuremath{\frac{\ensuremath{\chi_{2}^{2}}}{2}}\right]\right\rangle = -0.25068...$,
which gives: 
\begin{equation}
\left\langle log\left[\ensuremath{{P}(f_{j})}\right]\right\rangle =\left\langle log\left[I(f_{j})\right]\right\rangle +0.25068
\end{equation}

This method is suitable for evenly sampled data, which is not the
case for our data. Therefore, we first interpolated the data 
using a cubic spline interpolation method, and then applied the PSD method
on the interpolated light curves. Since the data has been manipulated, we 
tested our results via simulations.

\subsubsection{Simulations $\label{LCsim}$}

We tested the significance level of the obtained results via simulation.
The most used method is the Timmer \& K{\"o}nig method for light curve
simulations (\citealt{Timmer1995}). This method uses a Gaussian distribution
to simulate the light curves, which is not the case for the blazar light curves. The 
burst-like events in blazars deviate considerably
from such a distribution and are better described with a gamma or
lognormal distribution. This has been taken into account by \citealt{Emmanoulopoulos2013},
In this paper, we
simulated the light curves using the Emmanoulopoulos method using
the implementation
\footnote{http://ascl.net/1602.012} of (\citealt{Connolly2015}). 

The code calculates the probability density function (PDF) of a given
light curve. The estimated PDF is then fitted using  a combination of
a gamma- and a log-normal distribution. 
The code also
estimates the slope and the normalisation of the PSD, using a broken
power law model. However these estimates are not reported in this paper since
we used a simple power law fit for our PSD fits. 
Using the best fit parameters, and the slope and normalisation of the PSD, the code simulates
a number of light curves that have a PDF similar to the PDF
of the original light curve.

\subsubsection{PSD parameters test}

In order to test our results, a
raw fitting  on the original PSD is done using the function $P(f) = Nf^{-a}$,
(where $a$ is the slope and $N$ the normalisation). After setting the slope and the normalisation
as variable in the simulation algorithm (see section. $\ref{LCsim}$), an interval for each has
been defined with an appropriate increment. The next step is to sample
the simulated light curves at the same bin width as the observed
ones. For each combination of slope and normalisation, a total of two hundred light curve has 
been simulated. We calculated the PSD of each of these light curves, and calculated the mean
and standard deviation of the two hundred PSD as well. 

The final step is then to find the combination whose mean fits best the
original PSD and minimises the value of the Chi-square (\citealt{Uttley2002}): 
\begin{equation}
\chi_{dist}^{2}=\sum_{\nu=\nu_{min}}^{\nu_{max}}\frac{[\overline{P_{sim}(\nu)}-P_{obs}(\nu)]^{2}}{\Delta\overline{P_{sim}(\nu)^{2}}}
\end{equation}
By finding the combination that has a minimum Chi-square value, the slope and 
normalisation of the PSD of a given light curve  are then estimated. A simple 
power-law fit to this PSD was used for the error estimation for simplicity reasons. 
However, this is not the best approach as the the errors could be significantly underestimated. 
A more appropriate statistical approach to determine the PSD slopes uncertainties for unevenly sampled 
data is described in \citealt{Uttley2002, Emmanoulopoulos2013}.

\subsubsection{PSD results}

The variability in AGN light curves  
resembles various types of noise \citep{Vaughan2003,Vaughan2005}.
The observed variability is a convolution
of residual measurement errors and a mixture of the source intrinsic
processes, which means that the signal is a result of different statistical
noise processes; but in addition, also possible systematic variations
(e.g.\ by physically and geometrically caused variability). The PSD
analysis offers the best and the most commonly used method to analyse
and investigate the nature of variability. For PSD slopes $a$ close to zero,
the variability  resembles white noise processes. If the slope
is between $-$1 and $-$2, 
the variability is  considered to be consistent with red noise processes \citep{Vaughan2003}. 

\begin{figure}
\center
\includegraphics[scale=0.34]{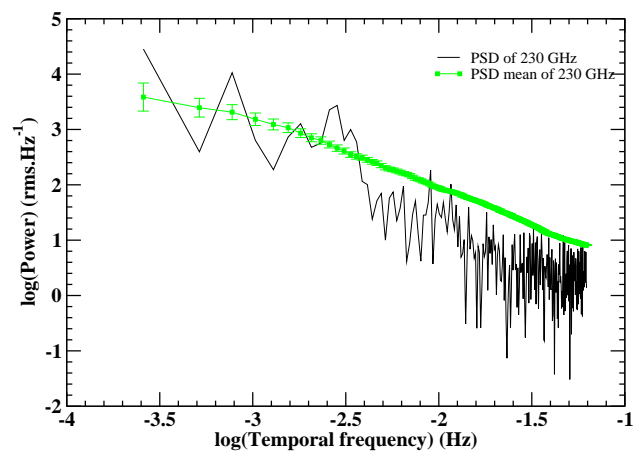}
\caption{Best-fitting average power spectrum (in green) superimposed
on top of the raw power spectrum at 230~GHz radio band (in black).}
\label{PSD_230}
\end{figure}

We applied the PSD method on all the observed light curves. Figure $\ref{PSD_230}$ shows an example of estimated PSD 
for the 230~GHz radio light curve; see Figs.~\ref{fig_app_4} and \ref{fig_app_5} for the PSD curves at other frequencies. The black curve is 
the raw PSD and the green fit curve shows the mean of 
the PSD values
of the two hundred simulated light curves of 230~GHz, with the slope (-1.6)
and normalisation (0.005) constant combination that minimised the Chi-square
value.

\begin{figure}
\center
\includegraphics[scale=0.34, trim = 0 0 0 2, clip]{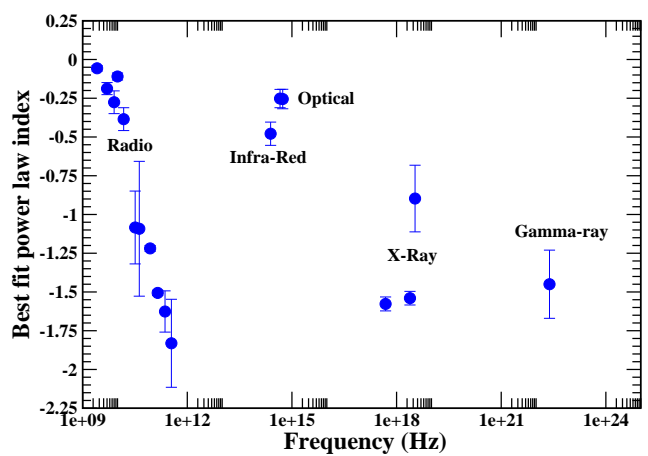}
\caption{The best fit power law index plotted as a function of the frequency}
\label{PSD_slope}
\end{figure}

In Fig.\ \ref{PSD_slope}, we plot the estimated PSD slopes as a function 
of frequency. 
A clear decreasing trend can be seen at radio frequencies. 
The PSD slope at radio bands gets steeper as we go higher in frequency suggesting 
a higher dominance of red noise at mm-radio bands compared to cm-bands. The PSD slopes get steeper as we go to higher frequencies,
which is consistent with the fact that at higher radio bands we are probing more and more rapid variations i.e.\ we are adding more power 
at higher PSD frequencies. As a consequence, the PSD slope should be flatter rather than steeper. However, this is not the case because of 
the following: at higher radio frequencies, we observe more flares because the emission region is optically thing. Moreover, there is a contribution
of long-term variations in addition to the flares observed in a given time window. 
We noticed that the PSD slopes at cm-radio bands ($a$) are consistent with white 
noise process, which could be well explained by opacity effects i.e.\ the 
emission region is opaque at these frequencies.   
The optical and IR light curves have PSD slopes that are consistent with white
noise process. 

The X-ray light curves at 2-10 keV band and 10-50 keV band have similar PSD slopes 
($a \sim -1.5$), which is also comparable to the slopes at mm-radio bands and at 
$\gamma$-ray energies; intensity variations at these energies are therefore well   consistent 
with red noise process. Variations at hard X-ray bands (14-192 keV) can also be described as
red noise process ($a \sim -1$); however, the  estimated PSD slope for this band 
significantly differs from the aforementioned bands. Different PSD slopes could be an 
indication of different processes responsible for the observed variations. 

\subsection{Cross-correlation analysis } \label{DCF}
We used the Discrete Correlation Function method \citep{Edelson1988} to investigate 
the correlation and possible
time lags between two light curves. The first step is to calculate 
the unbinned discrete correlation function (UDCF) using
the given time series for a time lag $\Delta t_{ij}=t_{j}-t_{i}$: 
\begin{equation}
UDCF_{ij}=\frac{(a_{i}-\bar{a})(b_{j}-\bar{b)}}{\sqrt{\sigma_{a}^{2}\sigma_{b}^{2}}}
\end{equation}
where $a_i$ and $b_j$ are the individual data points in two different
time series, with $\bar{a}$ and $\bar{b}$ being the respective mean
value of the time series, and $\sigma_a ^2$ and $\sigma_b ^2$ their
respective variance. 
The next step is to bin the UDCF in time. The DCF value for each bin is given as:
\begin{equation}
DCF(\tau)=\frac{1}{M}\underset{ij}{\sum}UDCF_{ij}(\tau)
\end{equation}
\noindent
where $M$ is the number of data points in each bin. We estimated
the error on the DCF using: 
\begin{equation}
\sigma_{DCF}(\tau)=\frac{1}{M-1}\sqrt{\sum[UDCF_{ij}(\tau)-DCF(\tau)]^{2}}
\end{equation}
\noindent
 This represents the standard deviation of the $UDCF_{ij}$
values within each bin. The $DCF$ value and its error of each interval,
is associated to the centre of the bin.
The profile of the DCF
curves obtained in this paper can be approximated by a Gaussian distribution.
As a conservative approach, the FWHM of the fitted Gaussian function is used as an error on the
estimated time lag. 

To test the significance  of the results
obtained with the DCF method, we simulated 2000 light
curves  as it is described in section $\ref{LCsim}$.
The simulated light curves have the same sampling, length, and PSD as the
original light curve. We cross-correlated the simulated light curves
with  the observed light curves in different
energy bands. We compared the DCF of the simulated light curves to
obtain the 95\% confidence level at a given time lag and generated
the confidence level curve\footnote{DCF peaks above the 95\%  confidence level are considered to be significant}. In the following section, we will discuss
in details the correlation between the broadband light curves.

\subsubsection{Radio--radio correlation}
\begin{figure}[!t]
\noindent \begin{centering}
\includegraphics[width=8cm,height=11cm,keepaspectratio]{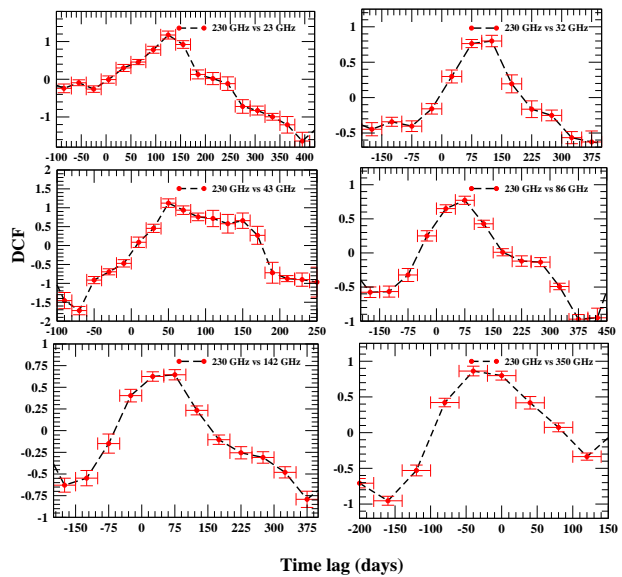}
\par\end{centering}
\caption{ Plot of the cross-correlation functions
of the 230~GHz data vs. the data at 23, 32, 43, 86, 142, and 350~GHz. }
$\label{RadioDCF}$
\end{figure}

\begin{figure}
 \begin{centering}
\includegraphics[scale=0.34,trim= 1 1 1 1.5, clip=true]{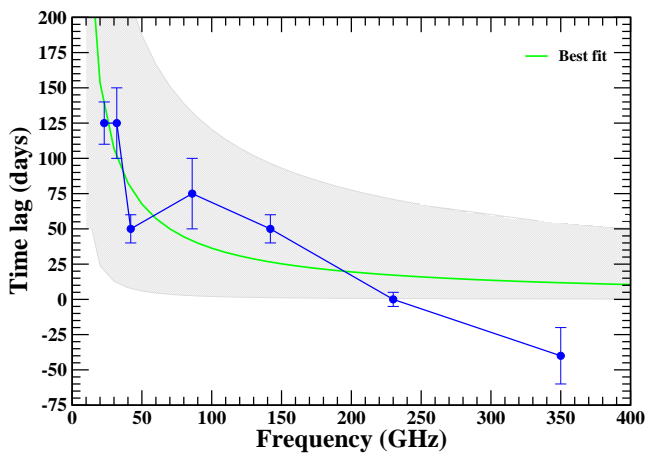}
\end{centering}
\caption{Time lag relative to the curve at 230\,GHz plotted as a function of radio frequencies (in blue). The
green curve represent the best power-law fit of the data points, with
a slope of $k_r$=1.12$\pm$0.46. The grey shaded area represents the error
interval on the fit. }
$\label{radio_TL}$ 
\end{figure}

As we have noticed in the previous section, the variability is more
pronounced at higher radio frequencies. Apparently, the flares at
high radio bands lead those at lower energies. To quantify
this, we used the DCF method and correlated every observed light curve
in the radio bands with the light curve at 230~GHz which is used as
a reference frequency because of its high cadence. We also used
 the 230~GHz light curve to check for  correlations
with the optical, X-ray, and $\gamma$-ray bands.  The
cross-correlation functions at radio bands are shown in Figs.~$\ref{RadioDCF}$ and \ref{fig_app_1}.
We found a correlation between 230~GHz and all the light curves at
frequencies between 23 and 350~GHz. For the light curves at frequencies
below 23~GHz, no significant correlation is noticed, which is to be expected
as the flaring  activity at low frequencies is barely visible. From
the DCF analysis, we derive the time lags, which we summarise in 
Table $\ref{Table}$  and Fig. $\ref{radio_TL}$. The decreasing
time lag with increasing frequency and the negative time lag for the
230 and 350~GHz correlation curve can be expected for the shock-induced flares \citep{Marscher1985,Fromm2011,Rani2013a}. 

\begin{table}[htbp]
\caption{Measured  time lags at radio frequencies as derived from DCF}
\label{Table}
\noindent \centering{}%
\begin{tabular}{cc}
\hline 
230~GHz vs.  & Time lag \\
             & (days)\\
\hline 
23~GHz & 125$\pm$15\\
32~GHz & 125$\pm$25\\
43~GHz & 50$\pm$10\\
86~GHz & 75$\pm$25\\
142~GHz & 35$\pm$15\\
230~GHz & 0$\pm$5\\
350~GHz & -(40$\pm$20)\\
\hline 
\end{tabular}
\\
The positive time lag indicates that the flaring activity at 230~GHz band is leading and vice-versa.
\end{table}

Figure $\ref{radio_TL}$ shows the estimated time lag between
the 230~GHz and the other radio light curves, plotted as a function
of frequency. The decreasing trend can be described by a power-law of the form $f(\nu)=N\nu^{-\frac{1}{k_{r}}}$ (\citealt{Kudryavtseva2011}).
The power-law  fitting gives $k_r$=(1.12$\pm$0.46), and the grey  shaded
area represents the error interval on the fit. A value $k_r$$\sim$1
suggests equipartition between the magnetic field and the
electron  energy density in the emission region (\citealt{Lobanov1998}).
The lack of correlation between 230~GHz and the lower radio frequencies
(below 23 GHz) could be due to synchrotron-self absorption,
where opacity effects dominate.

The estimated time lag could be translated into separation of the two emission regions using
(\citealt{Fuhrmann2014}):
\begin{equation}
\Delta r_{\gamma,r}=\frac{\beta_{app}c\tau_{\gamma,r}}{sin\theta}\label{eq.TL.DCF}
\end{equation}
where $\beta_{app}$ is the apparent speed, c is the speed
of light, $\theta$ is the viewing angle, and $\tau_{\gamma,r}$ is
the obtained time lag. The apparent speed $\beta_{app}$ has values
ranging from 5 to 15 for its different component (\citealp{Savolainen2006, Lister2013}).
In this paper, we adopt an averaged $\beta_{app}$ value of 7,
and for the viewing angle $\theta$=$9.8^{\circ}$ \citep{Savolainen2006}.

A recent jet kinematics study suggests that the location of the 7~mm (43~GHz) VLBI core is 
at a distance of 4--7~pc from the jet apex using core-shift measurements (Lisakov et al. 2016,
in prep). The correlation between 230~GHz and 43~GHz showed a time
lag of (50$\pm$10) days. This translates, by applying eq. $\ref{eq.TL.DCF}$,
to a relative separation of (1.7$\pm$0.3)~pc. Therefore, we estimate
that the 230~GHz emission region is at (5.3$\pm$0.3)~pc from the jet apex.

\subsubsection{Radio-optical correlation}
\begin{figure}
\begin{centering}
\includegraphics[scale=0.34, trim = 0 0 0 2, clip]{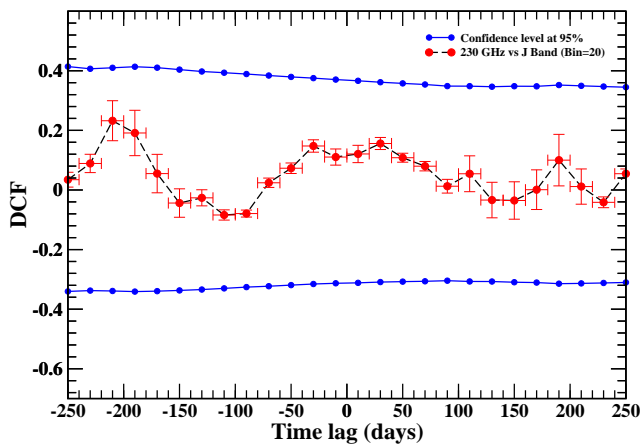}
\end{centering}
\caption{DCF analysis curve for 230~GHz vs. J passband. The blue curve
represents the 95\% confidence level.}
$\label{DCF_radio/jband}$
\end{figure}

Unlike the radio light curves, the
light curves at optical and IR passbands do not show any significant
variability. Therefore, we   cannot detect
any significant correlation between radio and optical/IR light curves.
In Fig. $\ref{DCF_radio/jband}$, we show the cross-correlation analysis
curve between radio (230~GHz) and J passband. 

\subsubsection{Radio -- X-ray correlation}
\begin{figure}
\begin{centering}
\includegraphics[scale=0.34]{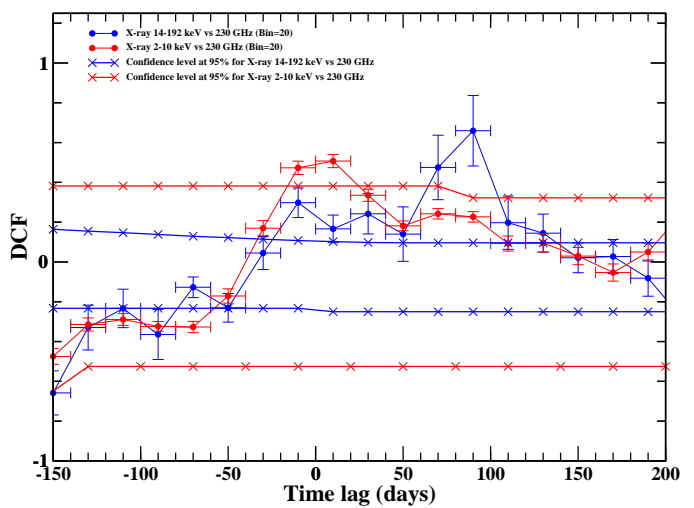}
\end{centering}
\caption{DCF curves of: X-ray at 14-192 keV vs. 230~GHz (blue) and X-ray at 2-10 keV vs. 230~GHz (red).}
$\label{DCF_X-ray/radio}$
\end{figure}

Similar fractional variability amplitudes and PSD slopes for mm-radio band and X-ray 
light curves suggest a possible correlation between the two. 
To quantify this, we applied the DCF analysis method and
correlated the  X-ray light curves at 14-192~keV and at 2-10~keV with the radio light curve
at 230~GHz. The DCF curves are shown in Fig.
$\ref{DCF_X-ray/radio}$.  The peaks of
the DCF curves are above the 95\% confidence level. This implies that
the X-ray light curves at different bands correlate significantly
with the radio light curve at 230~GHz. We notice a peak at a time
lag of (90$\pm$10) days for the correlation curve between X-ray 14-192
keV and 230~GHz, with a coefficient of (0.7$\pm$0.2) with the X-ray emission leading the radio emission.
This time lag translates to 3.1$\pm$0.4~pc relative distance, by
applying eq. $\ref{eq.TL.DCF}$.
For the DCF curve between X-ray 2-10 keV and 230~GHz, we
measured a peak at a time lag of (10$\pm$10) days with the X-ray emission leading the radio emission and a correlation
coefficient of (0.5$\pm$0.3)  giving a relative
distance of (0.4$\pm$0.4)~pc.  Similar results have been found for the correlation between 10-50~keV and 230~GHz, implying that the two emission regions
at X-ray energies 2--10 and 10--50\,keV are co-spacial.

\subsubsection{Radio -- gamma-ray correlations}
\begin{figure}
\begin{centering}
\includegraphics[scale=0.34]{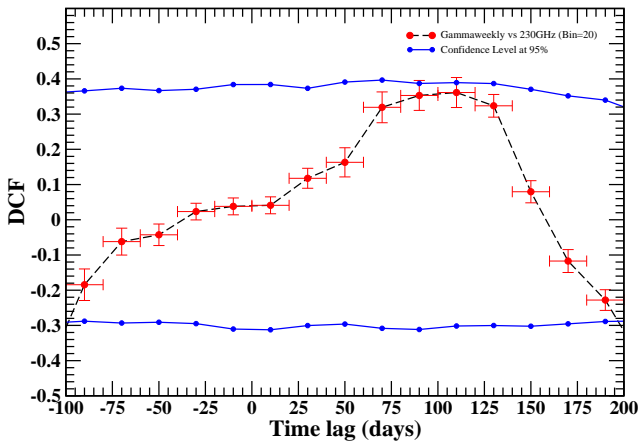}
\end{centering}
\caption{DCF curve of $\gamma$-ray and 230~GHz light curves - for the full data.
The blue lines represent the 95\% significance levels.}
$\label{DCF_gamma/radio_full}$
\end{figure}

As mentioned in section $\ref{analysis}$, there is an apparent correlation between the 
mm-radio and $\gamma$-ray light curves. To quantify this, we correlated the weekly sampled
$\gamma$-ray light curve with the radio light curve at 230~GHz. Figure
$\ref{DCF_gamma/radio_full}$ shows the DCF curve between $\gamma$-rays
and 230~GHz light curve for the entire data set. The DCF curve shows
a broad and prominent peak at a time lag of 110$\pm$ 27 days, suggesting
that  the $\gamma$-rays lead the radio emission
by $\sim$110 days. 
However, the formally calculated significance
of the cross-correlation falls below 95\%, with regard to the whole
data set. Since both radio and $\gamma$-ray light curves consist of
many flares, the DCF analysis of the entire data set shows an average
behaviour of different variability features. This could be due to
the blending effect of the two flares $\mathcal{R}_{1}$ and $\mathcal{R}_{2}$
in the radio bands. However, when we restrict
the DCF analysis to the time ranges of the individual flares $\mathcal{R}_{1}$
and $\mathcal{R}_{2}$, we found a  significant correlation
between the two. 

For the first flare, the DCF curve shows
a peak at 120$\pm$25~days, confirming that the  $\gamma$-rays
lead the radio emission. For the second flare, the DCF curve has a peak at (95$\pm$16) days. This suggests that the $\gamma$-ray emission
is leading the radio emission,  placing,
the $\gamma$-ray emission region closer to the jet apex than the radio
emission regions. Using eq.  $\ref{eq.TL.DCF}$,
a time lag of between 90 and 120~days gives  an 
offset of $\Delta r_{\gamma,r} $ between 3.1 and 4.1~pc.

\begin{figure}
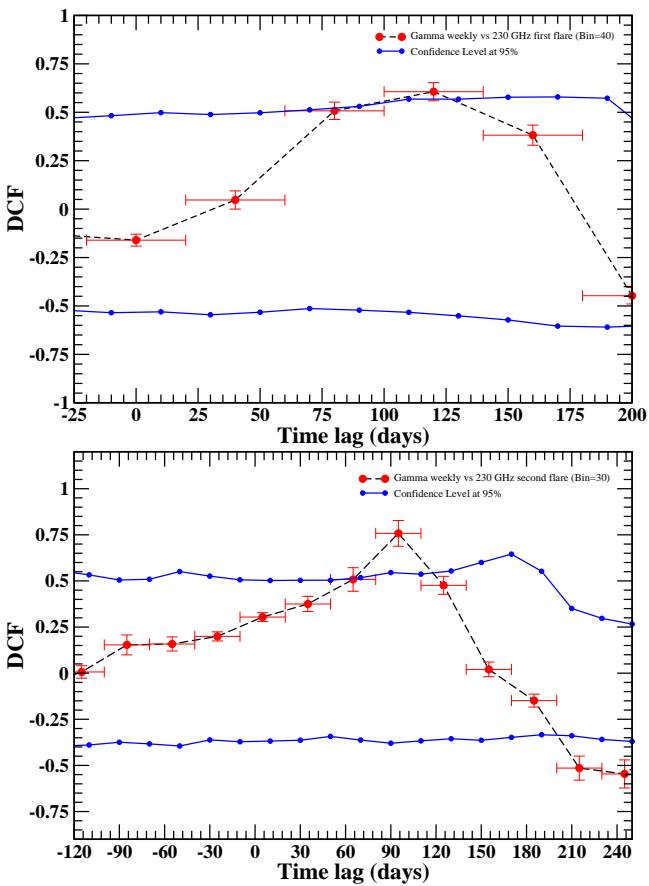

 \begin{centering}
\includegraphics[scale=0.34,trim=1 1 1 1.5, clip=true]{Gammaweeklyvs230_20_firstflare}
\includegraphics[scale=0.34,trim=1 1 1 1.5, clip=true]{Gammaweeklyvs230_20_secondflare}
\end{centering}
\caption{ DCF curve of $\gamma$-ray and 230~GHz during the first flare (top); 
DCF curve of $\gamma$-ray and 230~GHz during the second flare (bottom). The blue
lines represent the 95\% confidence levels.}
\end{figure}

\subsubsection{Optical - X-ray correlation}
The X-ray light curves show strong variability
across the observation period, which is not the case for the optical/IR
light curves. The absence of a prominent peak in the optical/IR light
curves, does not allow us to make any claims about a correlation
between the optical/IR and X-ray bands.  
A formal cross-correlation
analysis between the two bands does not show any peak.

\subsubsection{Optical/IR - Gamma-ray correlation}
As shown in  Fig. $\ref{LC_all}$, the
optical/IR light curves are not variable  during
the period of flaring activity in the GeV regime. The DCF analysis  shows that there is no correlation 
between the two bands.
Similar results are obtained for the correlation between $\gamma$-ray
and the optical R band, and between $\gamma$-ray and the infra-red
J band. 
This suggests different physical
mechanisms and causally disconnected emission regions for $\gamma$-rays
and optical frequencies. However, it is important to note the data gap in the optical/IR
light curve during the $\gamma$-ray flares, since from an observational point of view, a simultaneous
variability or delay cannot be discarded.

\begin{figure}
\begin{centering}
\includegraphics[scale=0.34, trim= 1 1 1 1.5,clip=true]{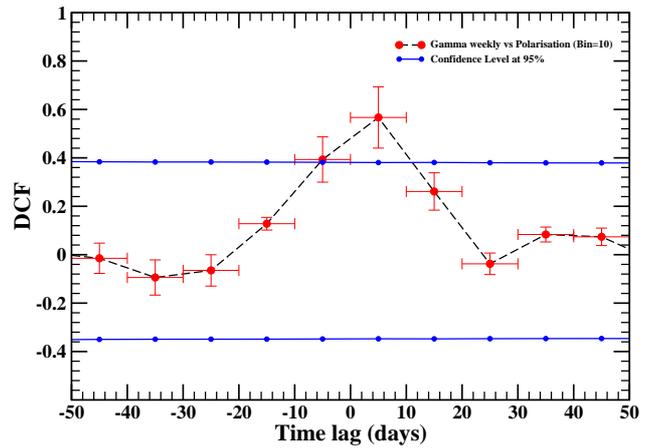}
\par\end{centering}
\caption{DCF curve of $\gamma$-ray light curve and optical percentage polarisation.
The blue lines represent the 95\% confidence levels.}
$\label{DCF_gamma/polarisation}$
\end{figure}

Although the polarisation observations in optical  band
are not as well sampled as the $\gamma$-ray light curve, there is
a hint of an apparent correlation between the two. For instance, we
noticed prominent variations in the polarisation curve during the
$\gamma$-ray flaring activity period. In addition to that, no variation
is apparent in the polarisation curve during the quiescent phase in
the GeV regime (see  Fig. $\ref{LC_all}$).
To quantify the apparent correlation, we applied the DCF analysis
method. The DCF curve is shown in  Fig. $\ref{DCF_gamma/polarisation}$. 
The DCF curve shows a peak at   a time lag of 5$\pm$5~days, which suggests 
a significant correlation between the two.  It is important to note that there is a data
gap in the percentage polarisation data between  MJD = 55000 and 55150 and between  MJD = 55400 and 55500. Therefore, it is
essential to test this correlation using simultaneous observations. 

\subsubsection{X-ray - X-ray correlations}
\begin{figure}
\begin{centering}
\includegraphics[scale=0.48, trim= 1 1 1 1.5,clip=true]{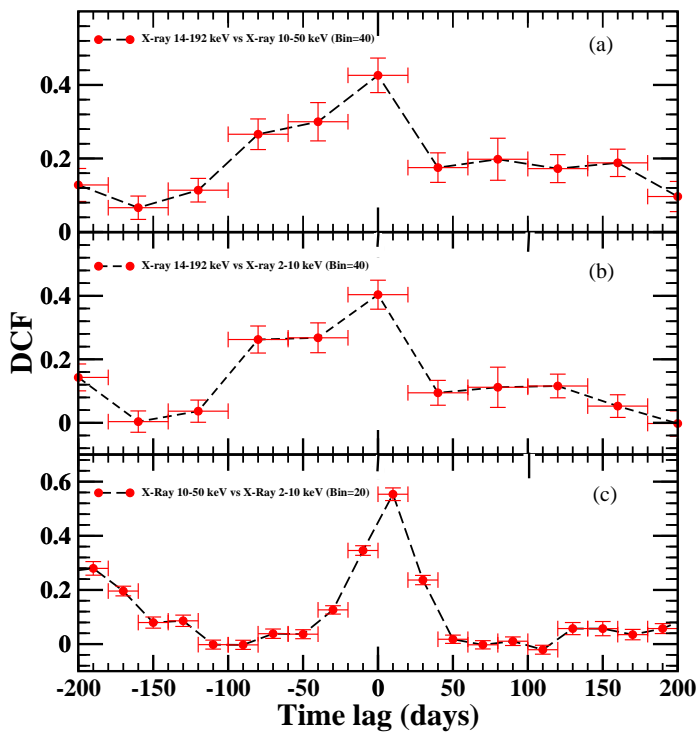}
\end{centering}
\caption{DCF curves of the different X-ray bands: 14-192 keV vs.\ 10-50
keV (a), 14-192 keV vs.\ 2-10 keV (b), and 10-50 keV vs.\ 2-10 keV (c). }
\label{X-raycorr}
\end{figure}

The X-ray light curves in the three different bands showed strong
and similar variability behaviour across the observation time. We
cross-correlated the X-ray light curve at 14-192 keV with the X-ray
light curves at 2-10 keV band and 10-50 keV band. The DCF results
 in Fig.\,\ref{X-raycorr}, show a correlation between the three X-ray bands. A peak close to 0 time lag is observed in the three DCF curves

\subsubsection{X-ray - $\gamma$-ray correlation}

\begin{figure}
\begin{centering}
\includegraphics[scale=0.34, trim= 1 1 1 1.5,clip=true]{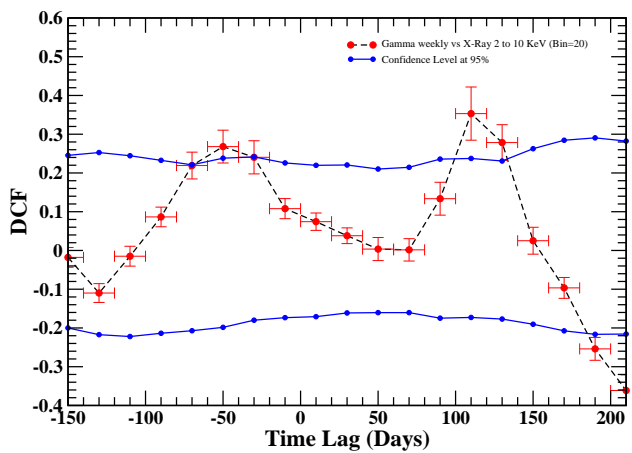}
\end{centering}
\caption{DCF curve of $\gamma$-ray and X-Ray 2-10 keV band. The blue lines represent
the 95\% confidence levels.}
$\label{DCF_gamma/X-ray}$
\end{figure}

The source exhibits prominent flaring activity at both X-ray and $\gamma$-ray
energies.  The DCF curve between $\gamma$-ray and the X-ray light
curve at 2-10 keV band is show in Fig. $\ref{DCF_gamma/X-ray}$. The DCF curve shows two peaks, at a time lag -(50$\pm$10) days and
(110$\pm$10) days. The positive time lag ($\sim$ 100~days) suggests that the $\gamma$-ray
is leading the X-ray emission. The negative time lag ($\sim$ -50~days) suggests that
the X-ray emission is leading the $\gamma$-ray emission. The estimated time
lags suggests the following two possible scenarios: 
\begin{enumerate}
\item Scenario 1: X-rays are leading the $\gamma$-ray emission with a time lag
of $-$(50$\pm$10) days, which translate to a de-projected distance
of (1.7$\pm$0.3)~pc using eq.
$\ref{eq.TL.DCF}$. The X-ray emission region is therefore closer
to the jet apex than the $\gamma$-rays emission region.

\item Scenario 2:  gamma-rays are leading
the X-ray emission with a time lag of 110$\pm$10~days, corresponding
to a distance of (3.8$\pm$0.3)~pc. The $\gamma$-ray emission region is then
closer to the jet apex than the X-rays emission region, or, in other
words, the $\gamma$-rays are further up stream, and the X-ray are emitted
from a region downstream the jet. 
\end{enumerate}

\section{Discussion}
$\label{discussion}$
Using the broadband variability study of the FSRQ 3C~273
observed between May 2008 and March 2012, we investigated the connection 
between flaring activity at GeV and lower energies. We also put constraint 
on the location and size of emission regions at different frequencies. The 
ultimate goal of the study is to provide a general physical scenario by which the observed variations of 
the source across several decades of frequencies can be put in a  consistent context.

\subsection{Radio emission}
The radio data collected between 2.6 to 350~GHz allow us to perform
a detailed study of flaring activity at radio bands. Two major outbursts
($\mathcal{R}_{1}$ and $\mathcal{R}_{2}$) were detected in the source
during the course of our observations. We found a significant correlation
between the multi-band radio data such that the
flaring activity at higher radio frequencies lead those at lower frequencies.
These results can be explained in the frame of the shock-in-jet-model (\citealt{Marscher1985}).
There was no pronounced flaring activity at radio bands below 23~GHz.
The fractional variability showed an increasing trend with increasing
radio frequencies. The PSD slopes suggests that the lower radio frequency variations 
are dominated by white noise processes. We then used the cross-correlation results 
between 43 and 230~GHz to locate the emission region at 230~GHz, which we found is about
(5.3$\pm$0.3)~pc from the jet apex.

\subsection{Optical/IR emission}
Compared to the other frequencies, the observed flaring activity is
significantly  lower at optical and IR
regime. The estimated fractional variability amplitude is lower than
13\%. In addition to that, the optical/IR light curves do not exhibit
a significant correlation with the observed light curves at other
frequencies  (which could also be due to the gap in the optical data). 
The power spectral density (PSD) slope results show a
power law index close to zero, which
is consistent with the slope of a white noise process. Spectral studies
done on the optical/IR emission showed that this emission, particularly
the excess emission at UV bands causing the blue bump, is dominated
by thermal emission from the accretion disk (\citealt{Shields1978,Soldi2008}).
Alternatively the presence of a X-ray
source or a hot corona could contribute significantly to the optical/UV
emission (\citealt{Collin-Souffrin1991,Haardt1994}).

\subsection{Gamma-ray emission}
The source 3C~273 showed a prominent flaring activity at GeV energies
during a period between July 2009 and April 2010. 
Our analysis suggests a significant correlation
between the $\gamma$-rays and the 230~GHz light curve, with a time lag
of (110$\pm$27) days, which translates to a relative separation between
$\gamma$-ray and 230~GHz emission region of around (3.8$\pm$0.9)~pc.
The radio correlation between 43 and 230~GHz puts the location the
230~GHz emission region at a distance of (5.3$\pm$0.3)~pc from the
jet apex. This places the $\gamma$-ray emission region at a distance
of (1.2$\pm$0.9)~pc from the jet apex. This result is in agreement
with the result found in \citealt{Rani2013} that constrained the
$\gamma$-ray emission region to < 1.6~pc from the jet apex. 

The fractional variability amplitude in
the radio bands is around 40\%, whereas, the $\gamma$-ray light curves
showed a fractional variability amplitude of around 100\%. The PSD
slope at mm-radio bands and $\gamma$-ray energies are however similar,
and are consistent with the slopes of a red noise process. Moreover,
the longest time scale of variability at high radio frequencies is
very similar to that at $\gamma$-ray energies, implying thus, a similar
size of the emission region for the two
energy ranges. The fastest variability time scale
for $\gamma$-rays frequency is 1.1 hour (\citealt{Rani2013}), suggesting
then, the presence of several very compact emission regions at $\gamma$-ray
energy bands, as it is expected in the multi-zone emission model suggested
by \citealt{Marscher2014}. 

The spectral changes reported in \citealt{Rani2013} suggests an external Compton mechanism
with seed photons from the  broad line region (BLR). 
The optical polarisation has
significantly increased during the $\gamma$-ray flaring activity, suggesting
thus a possible correlation between the $\gamma$-rays and the optical
polarisation.  This indicates that the non-thermal
emission component and the $\gamma$-ray emission are linked. 
A detailed broadband spectral energy distribution (SED) modeling  will 
shed more light on SSC (synchrotron self-Compton) and EC (external-Compton) 
contribution.

\begin{figure}
\includegraphics[scale=0.33, trim=20 50 1 50, clip=true]{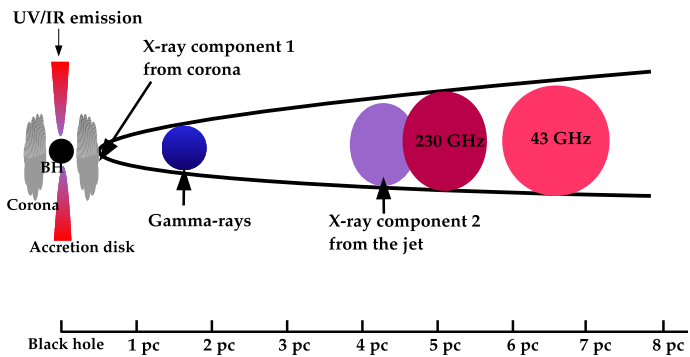}
\caption{Sketch representing the location of the emission regions
at different frequencies$\label{sketch}$. }
\end{figure}

\subsection{X-ray emission}
In comparison to a brief episode of flaring activity (July 2009 to
April 2010) followed by a quiescence phase at radio and $\gamma$-ray
frequencies, the X-ray regime was dominated by episodes of repeated
variability.  
The DCF curve between $\gamma$-rays and X-rays shows  two time
lags, one being negative of $-$(50$\pm$10)~days, and another
being positive with a time lag of 110$\pm$10~days. The DCF analysis suggests the presence of two possible components that are responsible for the X-ray emission. 
The negative
time lag suggests the X-ray emission is leading the $\gamma$-ray emission, and 
is located at a distance of 0.1 to 0.5~pc from the jet apex.  
The positive time lag suggests that the $\gamma$-ray emission
is leading the X-ray emission with a corresponding separation distance
of around (3.8$\pm$0.3)~pc. This component could be emitted from
the jet through inverse-Compton processes. An independent study of the X-ray spectrum 
also suggests the presence of two components. A component
coming from a hot corona or reflected off an accretion disk, accounting
for the presence of the weak iron line found in the X-ray spectrum
at 3-78 keV, and a second component most likely coming from the jet that starts
to dominate at 30-40 keV \citep{Madsen2015}.

The two component scenario for the X-ray emission is also confirmed by 
X-ray vs.\ radio correlation analysis. We noticed similar fractional variability 
amplitudes for the three X-ray bands and the mm-radio bands. The PSD slopes
are also comparable for the mm-radio bands and X-ray at 2-10 keV and 10-50
keV. The SF analysis showed the presence of a quasi-periodic variability
for both X-ray and mm-radio light curves. 
Correlation analysis 
suggests a time lag of 90$\pm$10~days  between 14-192 keV X-ray light curve 
and 230~GHz light curve, which translates to 3.1$\pm$0.4~pc placing
the location of the X-ray emission region at 0.5 to 2~pc from the jet apex. 
The DCF curve between X-rays at 2-10 keV
and 230~GHz shows a time lag of (10$\pm$10) days, which gives a relative
distance between the X-ray and radio emission regions of (0.4$\pm$0.4)~pc. 
This provides an estimate of the location of 2-10~keV X-ray emission 
at 4.7$\pm$0.4~pc from the jet apex. We obtained similar results for 
the intensity variations at 10-50~keV X-ray band.  

The longest time scale of variability for the X-ray light curves
is $\sim$230\,days. To estimate the size of the emission region,
we applied (\citealt{Rani2013}): 
\begin{equation}
R\leq\frac{c\times t_{var}\times\delta}{1+z}
\end{equation}
where $R$ is the size of the emission region, $t_{var}$ is the shortest
time scale of variability, $c$ is the speed of light, $\delta$ is
the Doppler factor of the source for which we used a value of 7 \citep{Savolainen2006}, and
$z$ is the red-shift of the source. For a time lag of $\sim$230\,days,
we obtained an emission region with a size of  $\sim$1.2~pc. The
size of this emission region is much bigger than the size of the corona,
which was suggested to be around 20-30\,$R_{S}$ (0.013-0.02~pc)  (\citealt{Reis2013}) where
$R_{S}=6.4\times10^{-4}$~pc is the Schwarzschild radius of the black
hole where the mass of supermassive black hole
of 3C~273 is around $6.6\times10^{9}\,\mathrm{M_{\odot}}$ (\citealt{Paltani2005}).
However, we noticed that for the X-ray light curves used in this study,
the shortest variability time scale is $\sim$ 5-10
days. This gives translates to a size of 0.025 -- 0.05~pc, 
which is comparable to the size of the corona. This implies,
that the long term variability is probably coming from the jet, whereas the faster variability seems to be originated closer to the black hole.

\subsection{Broadband correlation alignment}
The statistical analysis presented in this paper suggests a  causal connection between 
the observed broadband flaring activity in the source. The proposed scenario is presented in 
Fig. $\ref{sketch}$  where we marked the location of different emission regions. 
A moving shock/disturbance propagating downstream of the jet could first produce the 
$\gamma$-ray flares at a distance of $\sim$1.2~pc, and later could brighten the emission region 
at mm-radio bands. The mm-radio flares lead those at cm-radio bands.  Delayed emission at 
cm-radio bands is a clear indication of opacity effects due to synchrotron self-absorption. 
Our analysis suggests two possible components responsible for the X-ray emission.  One component 
of the X-ray emission is most likely coming from the hot corona or jet-apex. The second 
X-ray component seems to have non-thermal origin i.e.\ produced via inverse-Compton scattering 
of radio synchrotron photons at a distance of $\sim$4.7~pc from the jet-apex. Observations at 
optical/IR regime suggests a prominent thermal dominance either from the accretion disk or/and 
from the BLR region.

\section{Conclusion}
$\label{conclusion}$
In this paper, we presented the results
of our broadband variability study of the FSRQ 3C~273 observed for
a time period between May 2008 and March 2012. A detailed statistical
analysis was performed to constrain the location of the
emission region at different energy bands and to explore the connection
between the broadband flaring activity. 

The source went through a
series of rapid flares at the $\gamma$-rays which has been been accompanied
by flaring activity at other energies. We studied the broadband variability
behaviour of 3C~273, by comparing the general behaviour and the properties
of the light curves at different energy bands. 
Except for the light curves at cm-radio 
and optical/IR frequencies,  the observed broadband variations are found to be 
consistent with red-noise processes.  Absence of a pronounced flaring activity at 
at cm-radio bands could be due to synchrotron self-absorption, and at optical/IR bands, it 
could be because of the dominance of thermal emission. The optical/IR emission seems to be 
dominated by thermal emission from the disk and/or the BLR.

At radio bands, the fractional variability amplitude increases with increase in frequency, 
while the variability time scale decreases implying that at higher frequencies,
we are probing more compact emission regions. In addition to that the variability time scales are 
found to be very similar at different regimes.  The estimated 
fractional variability seems to follow a double hump structure as a function of frequency, which 
suggests that the largest variations are seen for the highest energy photons. 

The evolution of radio flares could be well 
explained in the frame of the shock-in-jet model. From the correlation analysis, we found that 
the 230~GHz radio flares are emitted at a distance between 5.3~pc from the jet apex. 
The $\gamma$-ray energies are emitted from
a compact region at a distance $\sim$ 1.2~pc from the jet apex, through
IC processes. There seem to be two components responsible
for the X-ray emission. The first component is located within a distance of 
0.1--0.5~pc from the  jet apex and most likely emitted either by the corona or regions 
closer to the jet apex.  The second component responsible for X-ray emission 
is located at a distance of $\sim$4.7~pc and is produced through up-scattering of 
radio synchrotron photons via IC processes. A broadband spectral energy density analysis can provide better constrains on
the different emission mechanisms that are taking place in the jet.

\begin{acknowledgements}
CC was supported for this research through a stipend from the International Max
Planck Research School (IMPRS) for Astronomy and Astrophysics at the Universities
of Bonn and Cologne. The authors would like to thank Dimitris Emmanoulopoulos,  Sam Connolly, and
Walter Max-Moerbeck for useful discussions on statistical analysis and for their
suggestions and comments on the paper draft. This research is partly based on observations
with the 100-m telescope of the MPIfR (Max-Planck-Institut für
Radioastronomie) at Effelsberg.
The ${{\it Fermi}}$-LAT Collaboration acknowledges generous ongoing
support from a number of agencies and institutes that have supported
both the development and the operation of the LAT as well as scientific
data analysis. These include the National Aeronautics and Space Administration
and the Department of Energy in the United States, the Commissariat
à l'Energie Atomique and the Centre National de la Recherche Scientifique
/ Institut National de Physique Nucléaire et de Physique des Particules
in France, the Agenzia Spaziale Italiana and the Istituto Nazionale
di Fisica Nucleare in Italy, the Ministry of Education, Culture, Sports,
Science and Technology (MEXT), High Energy Accelerator Research Organization
(KEK), and Japan Aerospace Exploration Agency (JAXA) in Japan, and
the K.\textasciitilde{}A.\textasciitilde{}Wallenberg Foundation, the
Swedish Research Council, and the Swedish National Space Board in
Sweden. Additional support for science analysis during the operations
phase is gratefully acknowledged from the Istituto Nazionale di Astrofisica
in Italy and the Centre National d'Études Spatiales in France. 

The optical/IR observations are provided by the Yale {\it Fermi}/SMARTS
project. Data from the Steward Observatory spectropolarimetric monitoring
project were used. This program is supported by {\it Fermi} Guest Investigator
grants NNX08AW56G, NNX09AU10G, NNX12AO93G, and NNX15AU81G.

The Submillimeter Array is a joint project between the Smithsonian
Astrophysical Observatory and the Academia Sinica Institute of Astronomy
and Astrophysics and is funded by the Smithsonian Institution and
the Academia Sinica.  This research
is partly based on observations carried out with the IRAM 30m telescope.
IRAM is supported by INSU/CNRS (France), MPG (Germany) and IGN (Spain).
\end{acknowledgements}


\begin{appendix}

\section{Structure function, cross-correlation analysis, and PSD plots}

\begin{figure}[h]
\includegraphics[scale=0.32]{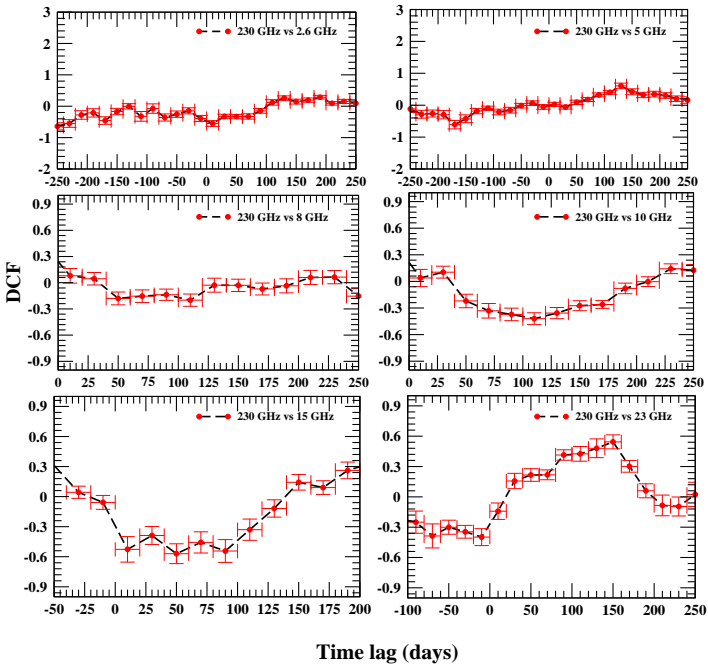}
\caption{Cross-correlation analysis curves at radio frequencies (230~GHz vs.\ 2.6, 5, 8, 10, 15, and 23~GHz). }
\label{fig_app_1}
\end{figure}

\begin{figure*}
\includegraphics[scale=0.18,trim=1 1 1 1.5, clip=true]{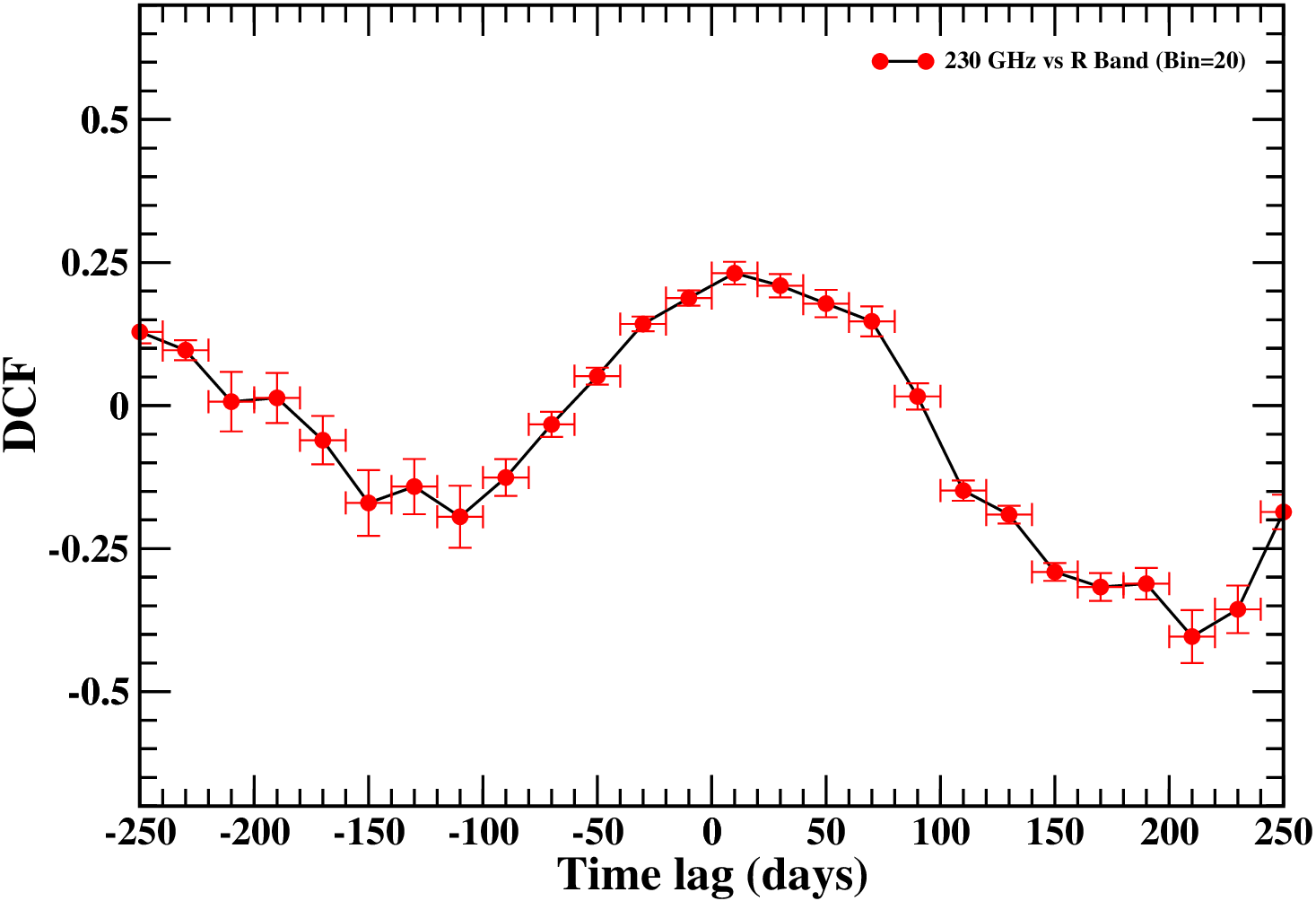}
\includegraphics[scale=0.18,trim=1 1 1 1.5, clip=true]{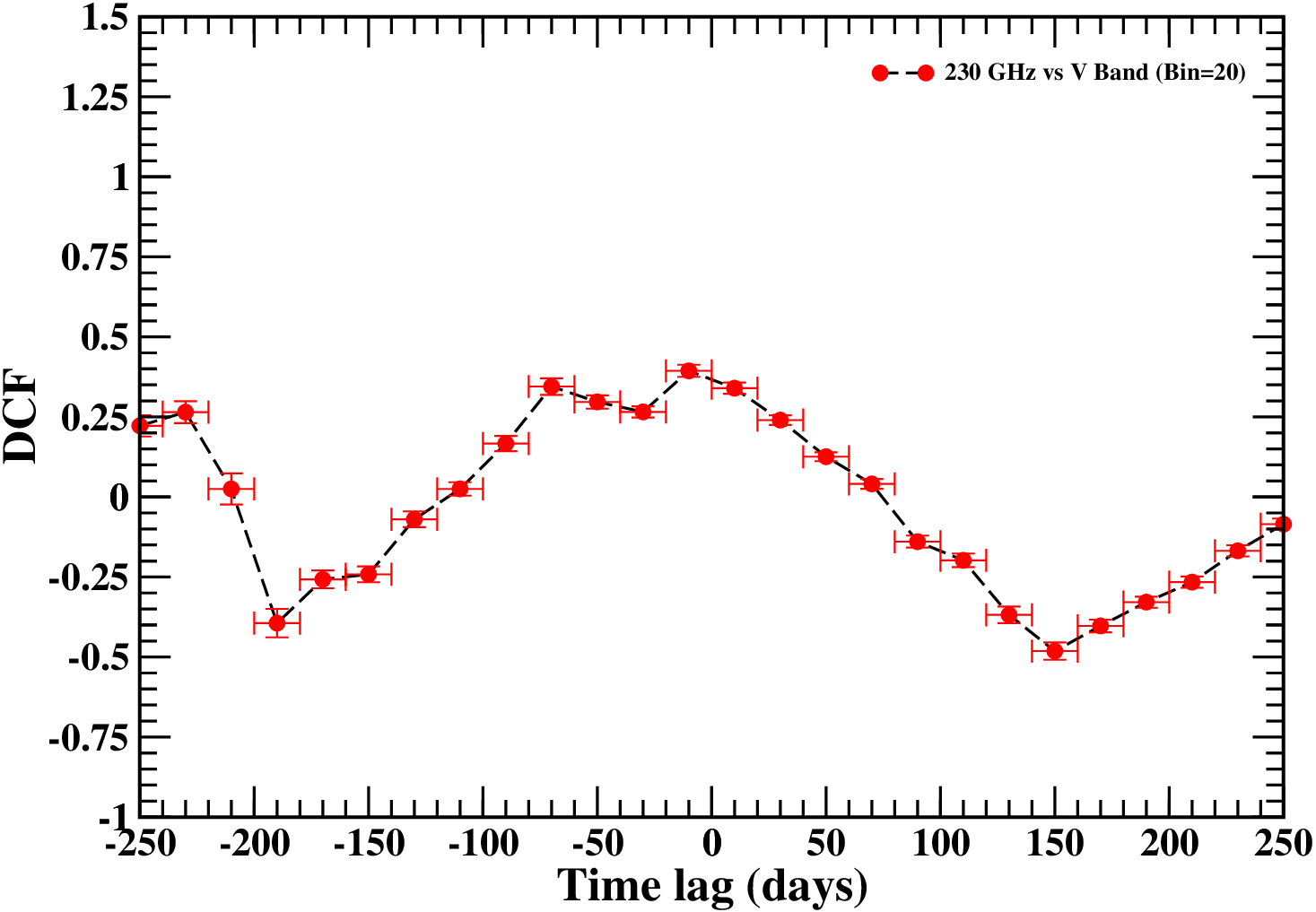}
\includegraphics[scale=0.18,trim=1 1 1 1.5, clip=true]{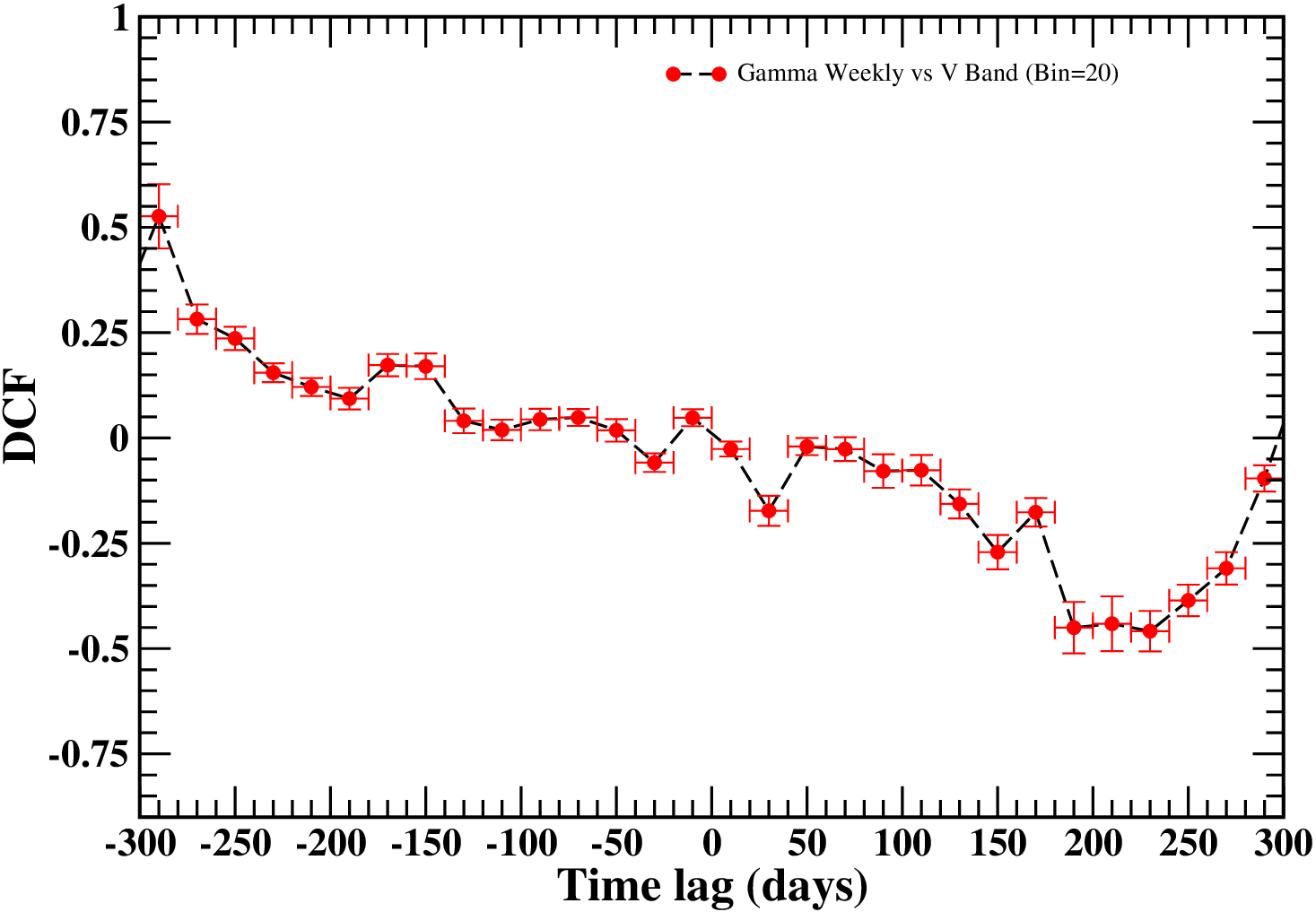}
\includegraphics[scale=0.18,trim=1 1 1 1.5, clip=true]{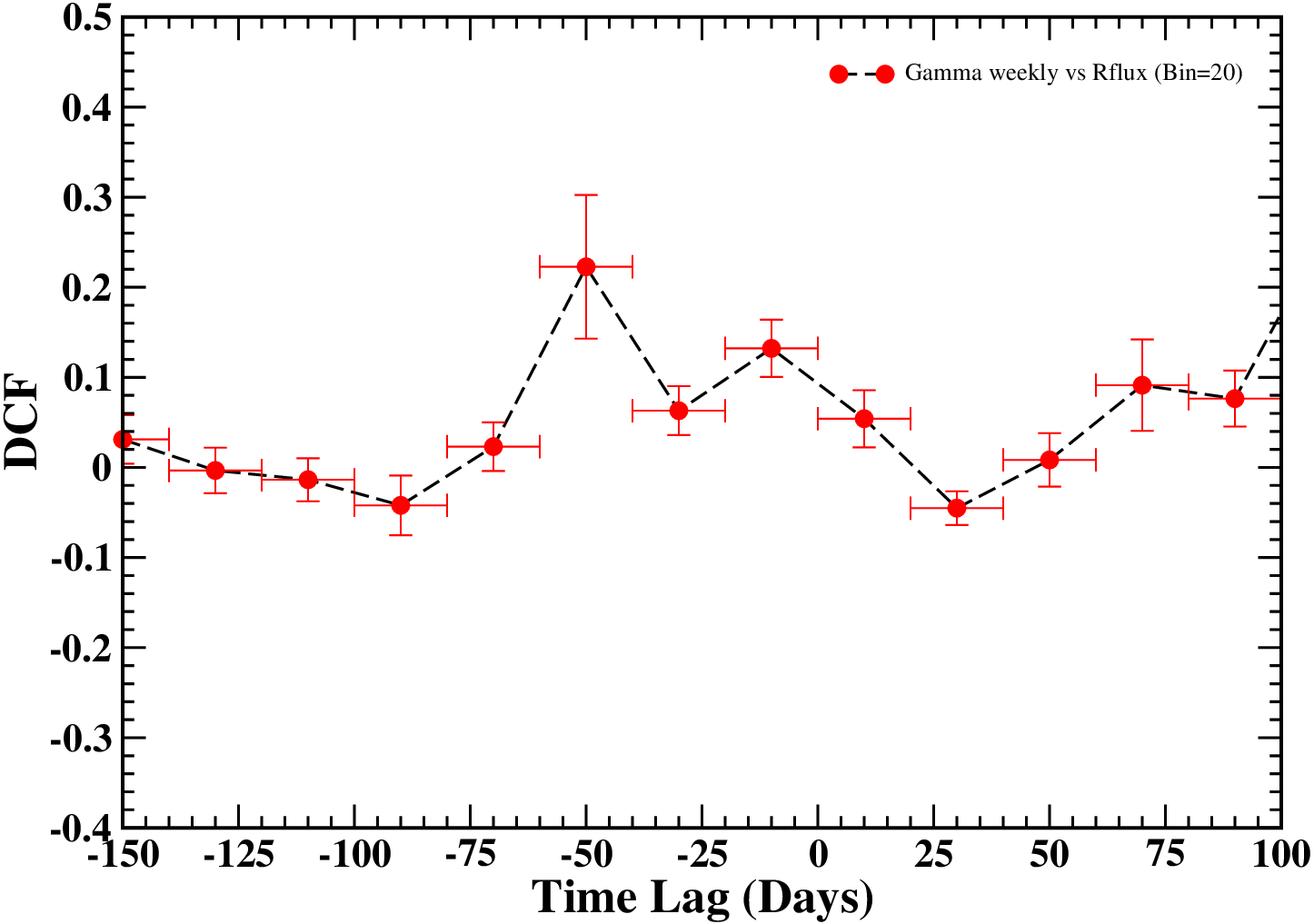}
\includegraphics[scale=0.18,trim=1 1 1 1.5, clip=true]{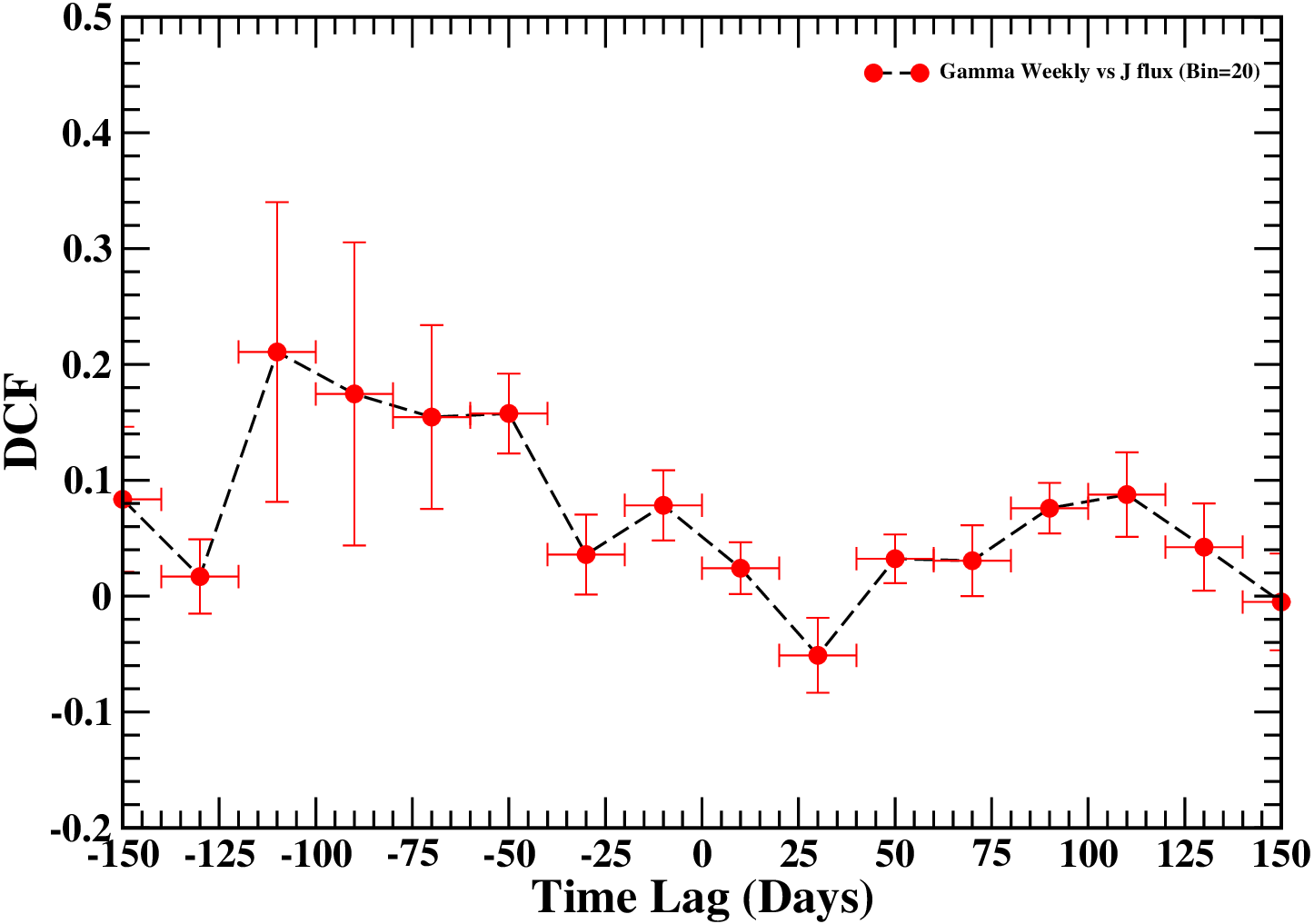}
\includegraphics[scale=0.18,trim=1 1 1 1.5, clip=true]{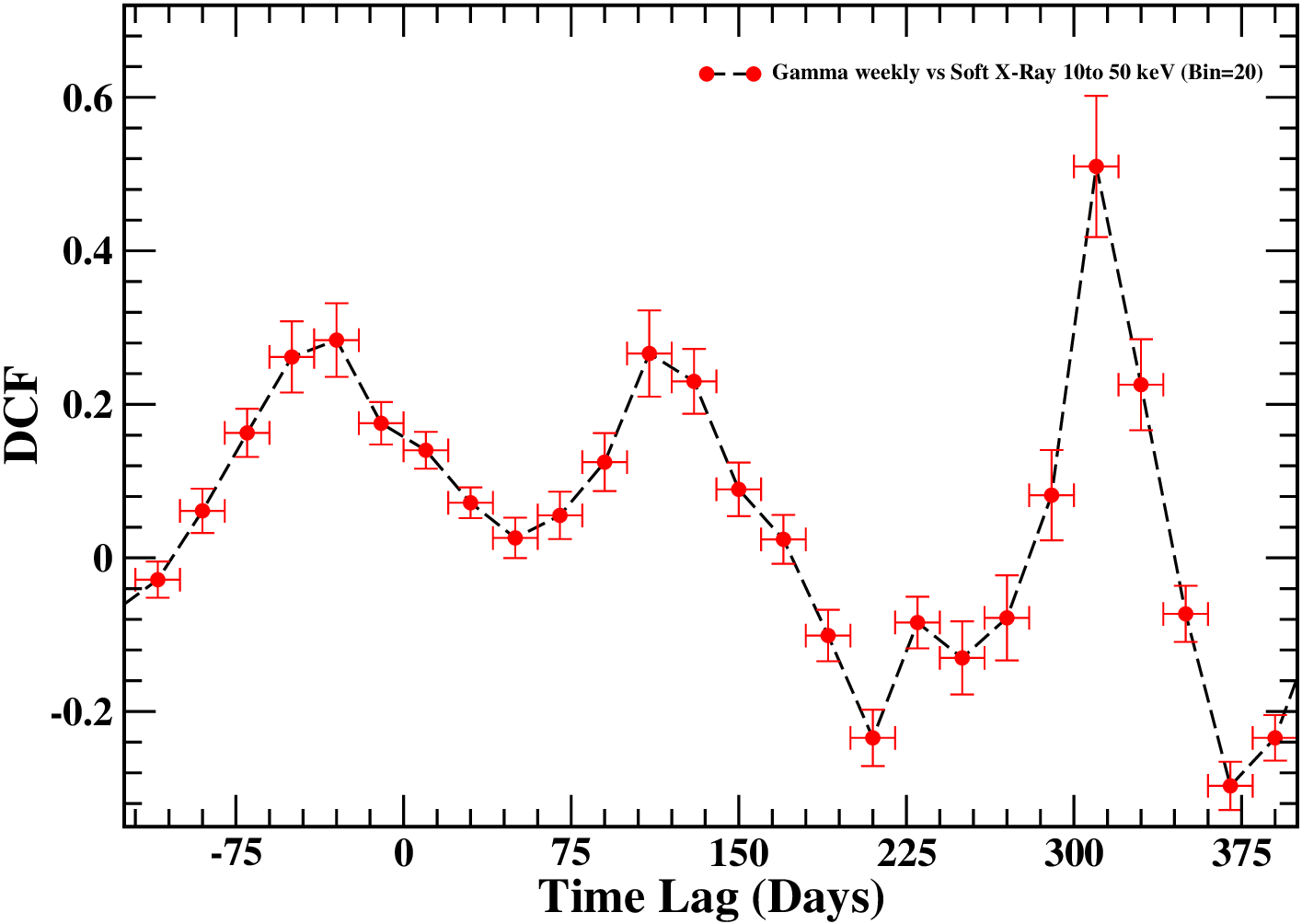}
\includegraphics[scale=0.18,trim=1 1 1 1.5, clip=true]{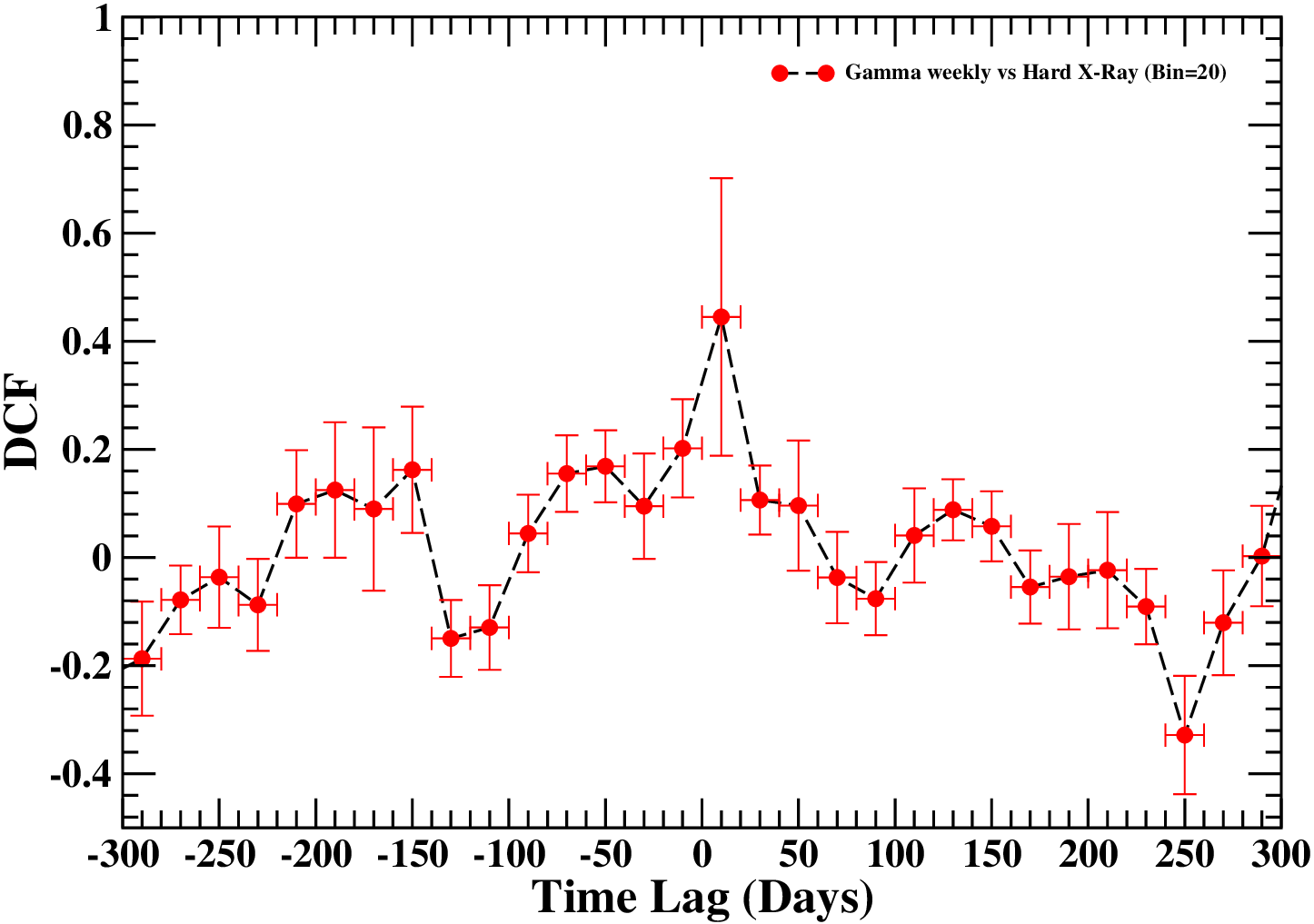}
\includegraphics[scale=0.18,trim=1 1 1 1.5, clip=true]{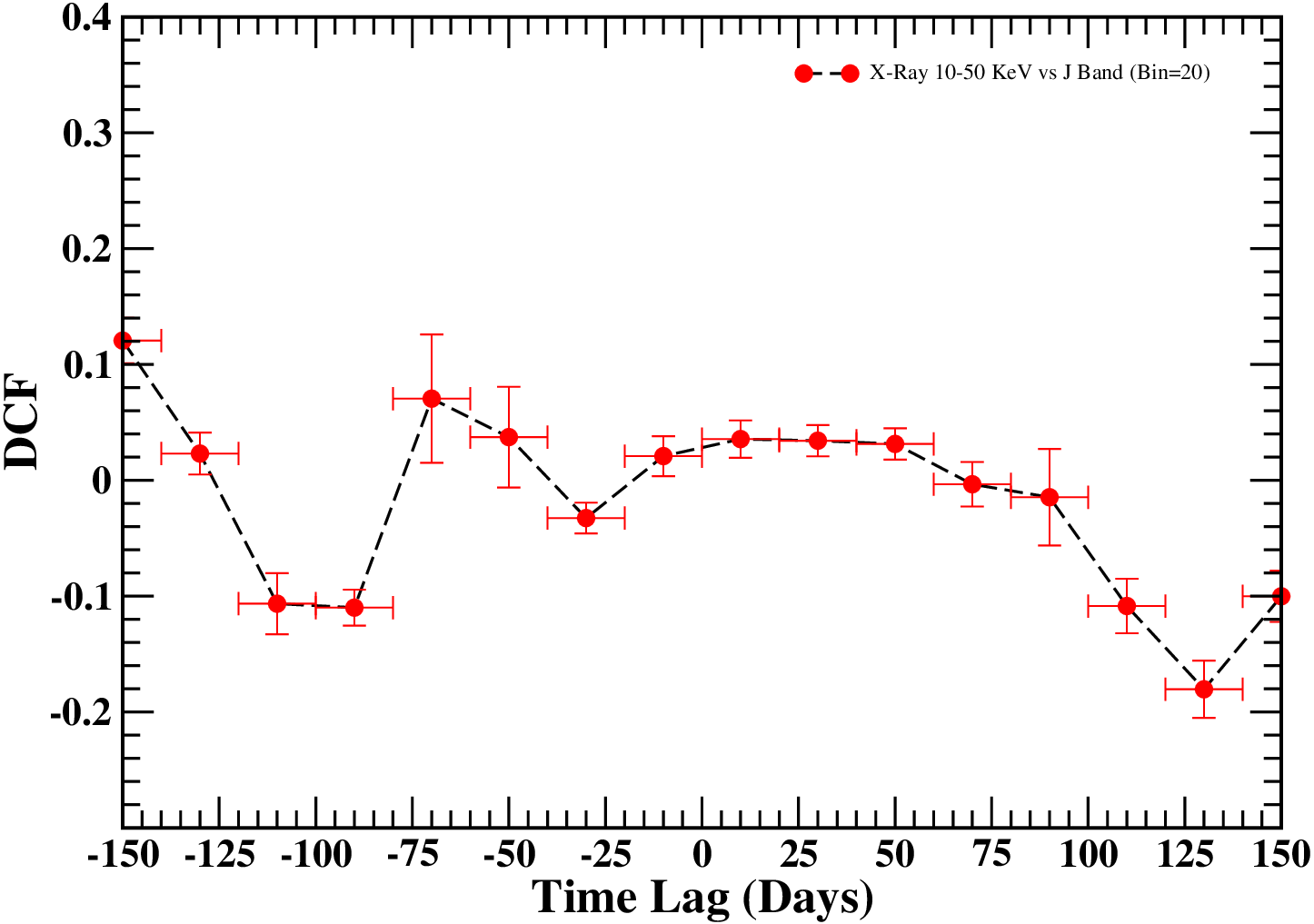}
\includegraphics[scale=0.18,trim=1 8 1 2.5, clip=true]{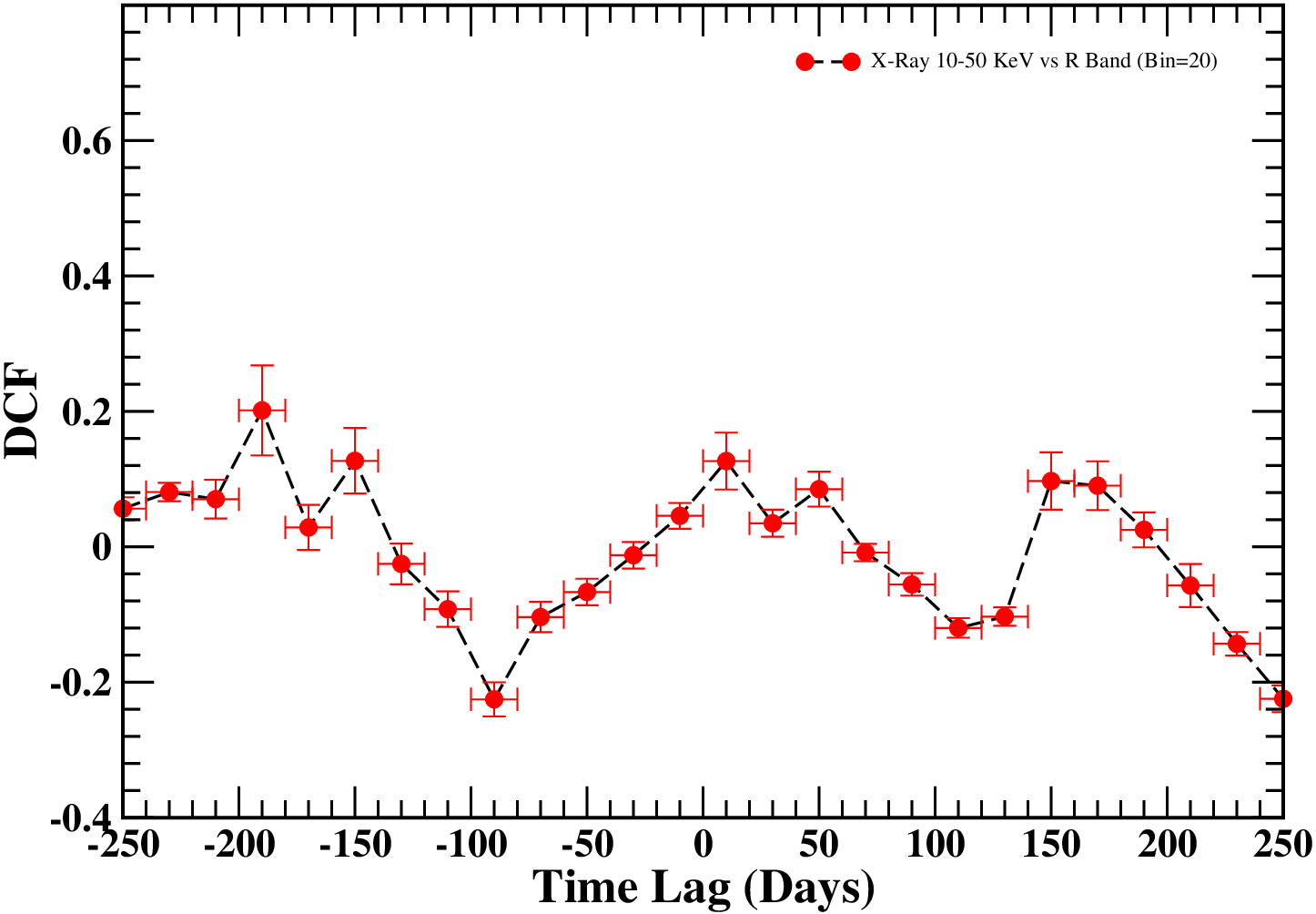}
\includegraphics[scale=0.18,trim=1 8 1 2.5, clip=true]{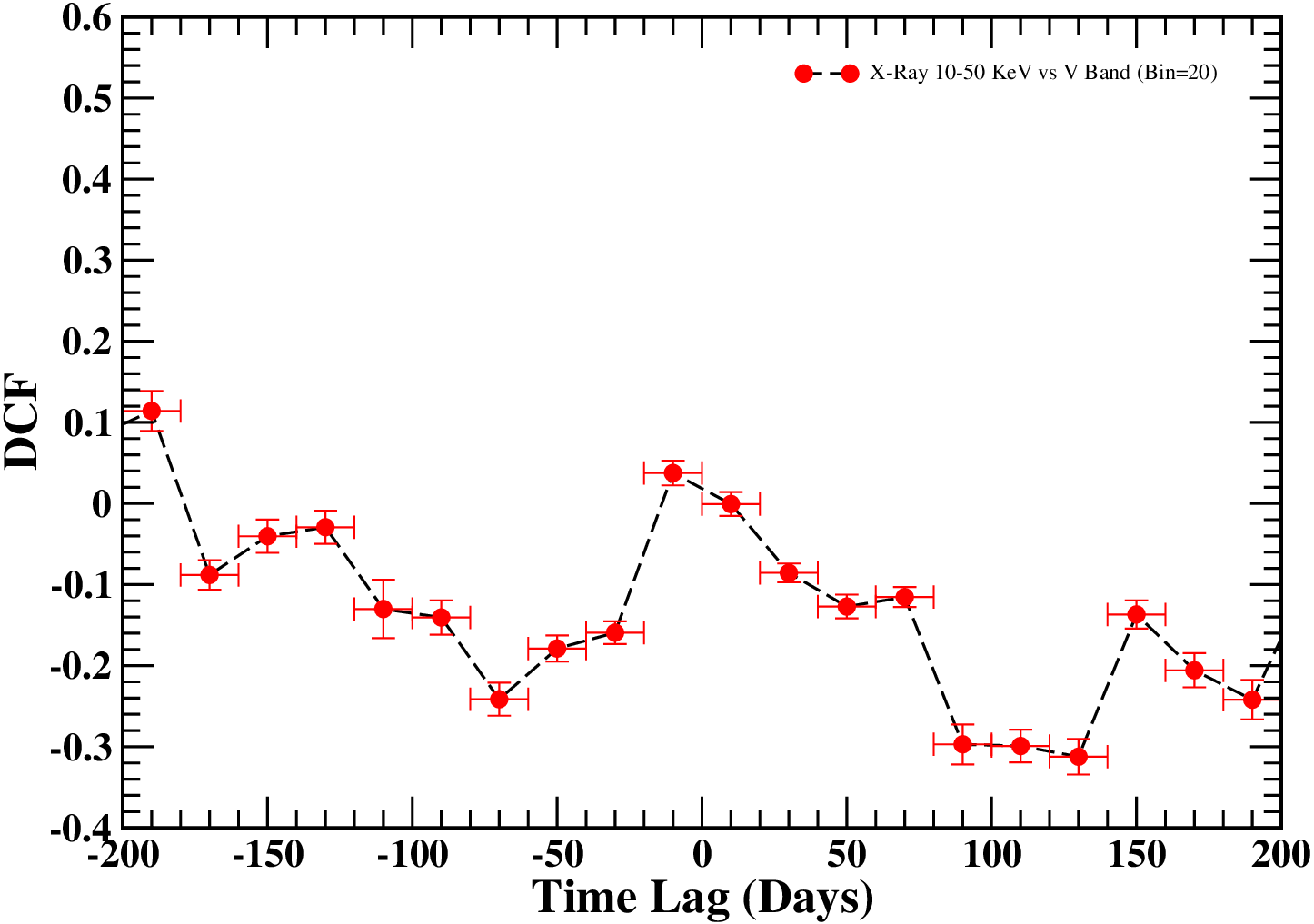}
\includegraphics[scale=0.18,trim=1 8 1 2.5, clip=true]{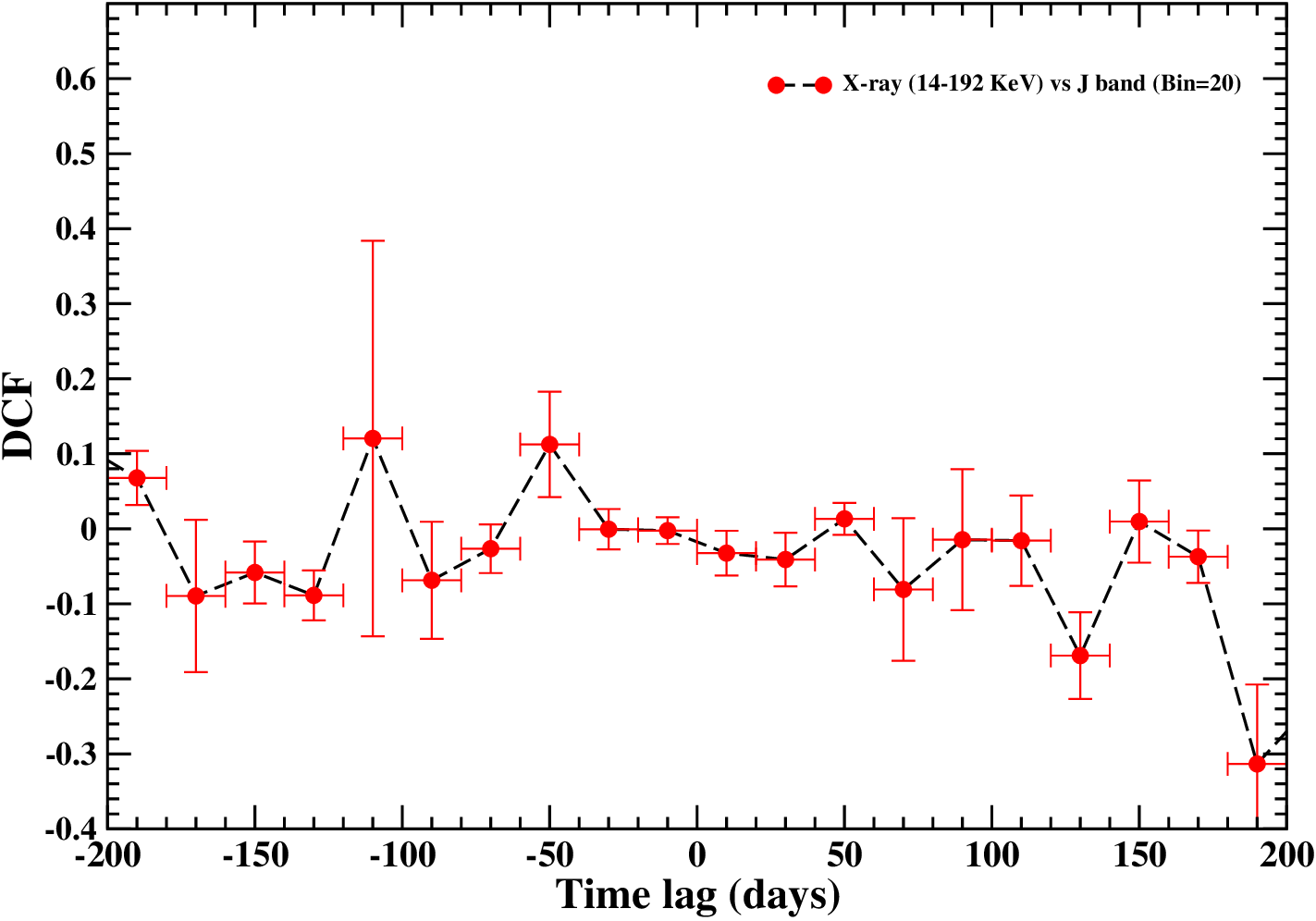}
\includegraphics[scale=0.18,trim=1 8 1 2.5, clip=true]{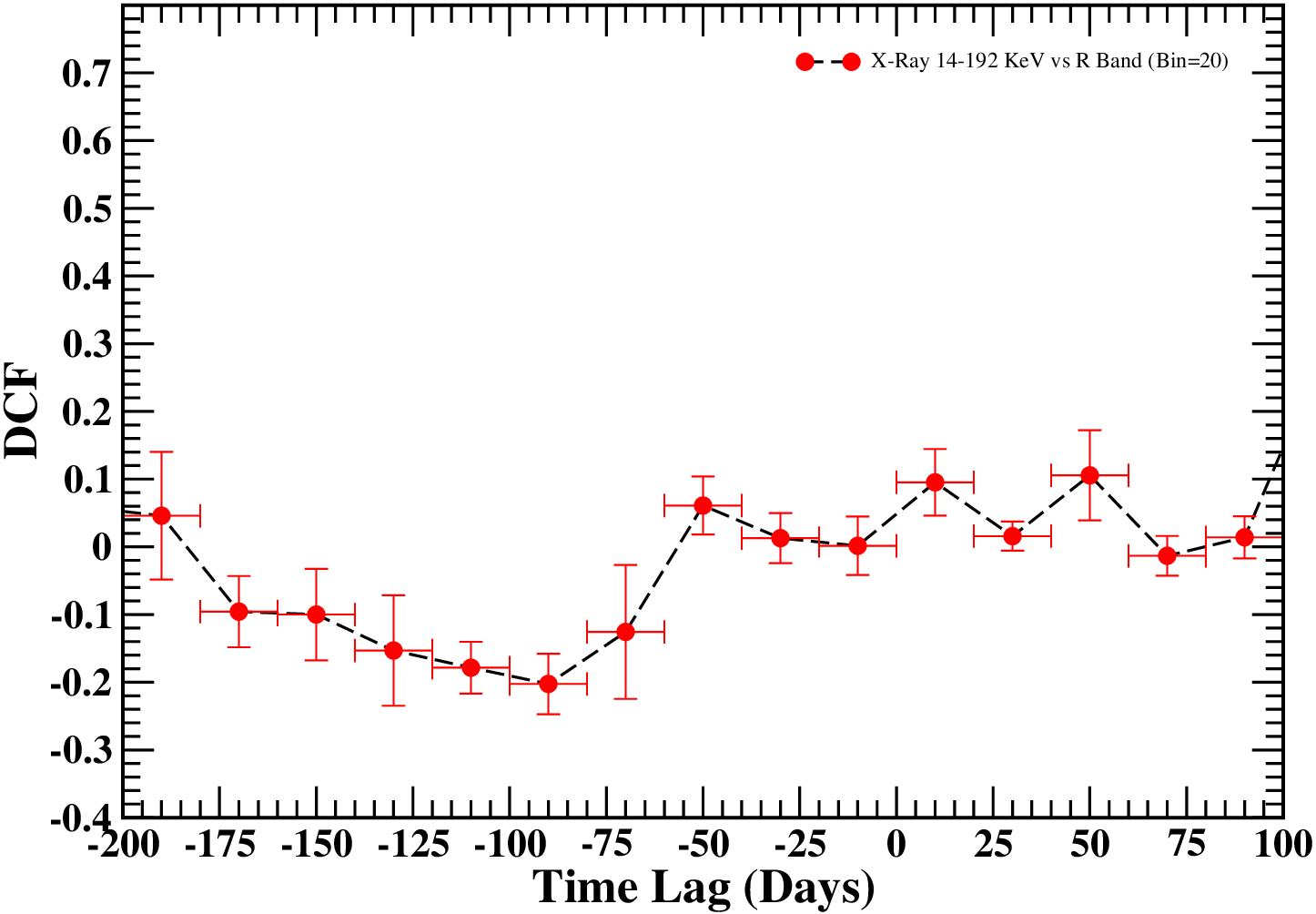}
\includegraphics[scale=0.18,trim=1 8 1 2.5, clip=true]{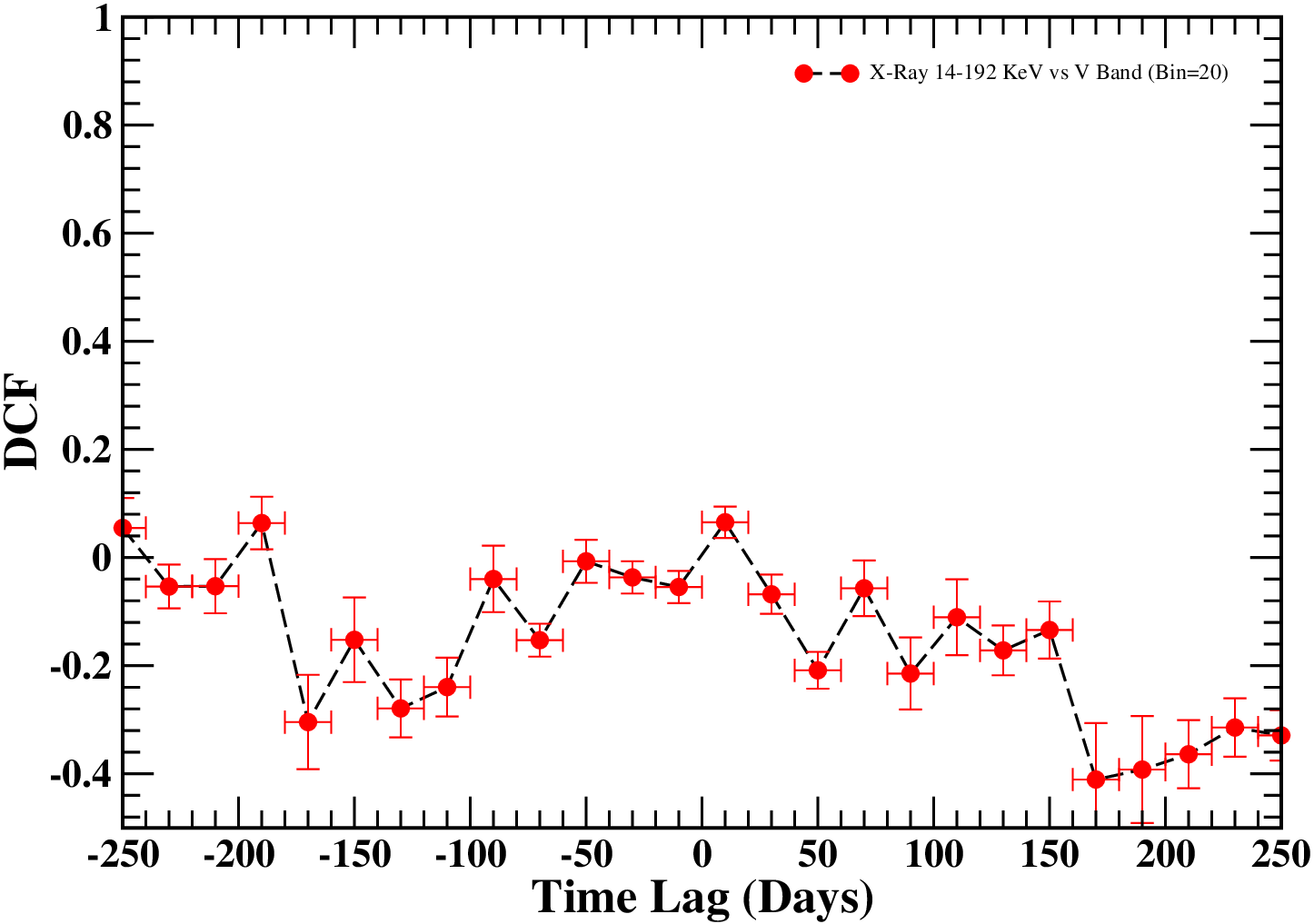}
\includegraphics[scale=0.18,trim=1 8 1 2.5, clip=true]{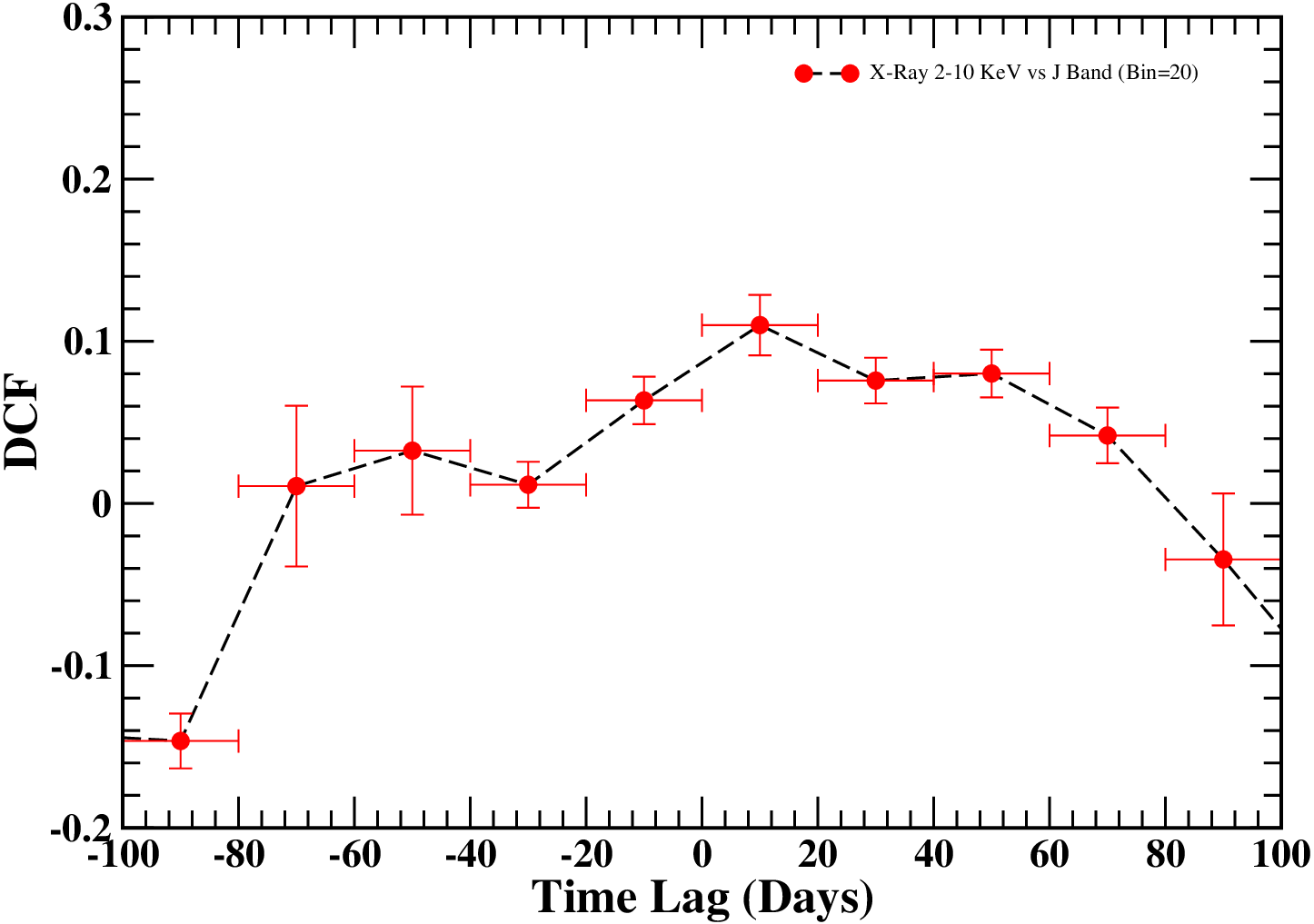}
\includegraphics[scale=0.18,trim=-30 8 1 2.5, clip=true]{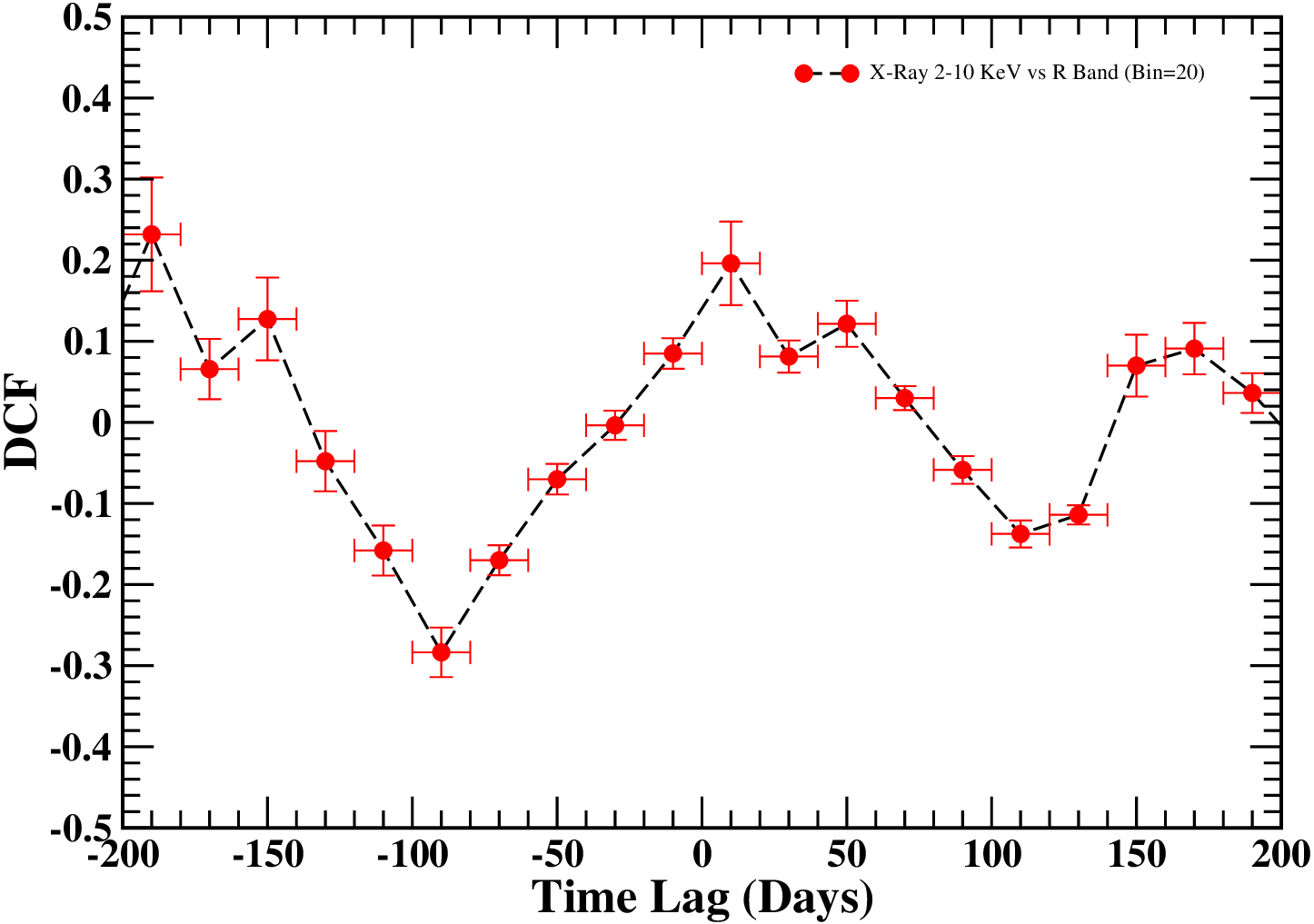}
\includegraphics[scale=0.18,trim=-50 8 1 2.5, clip=true]{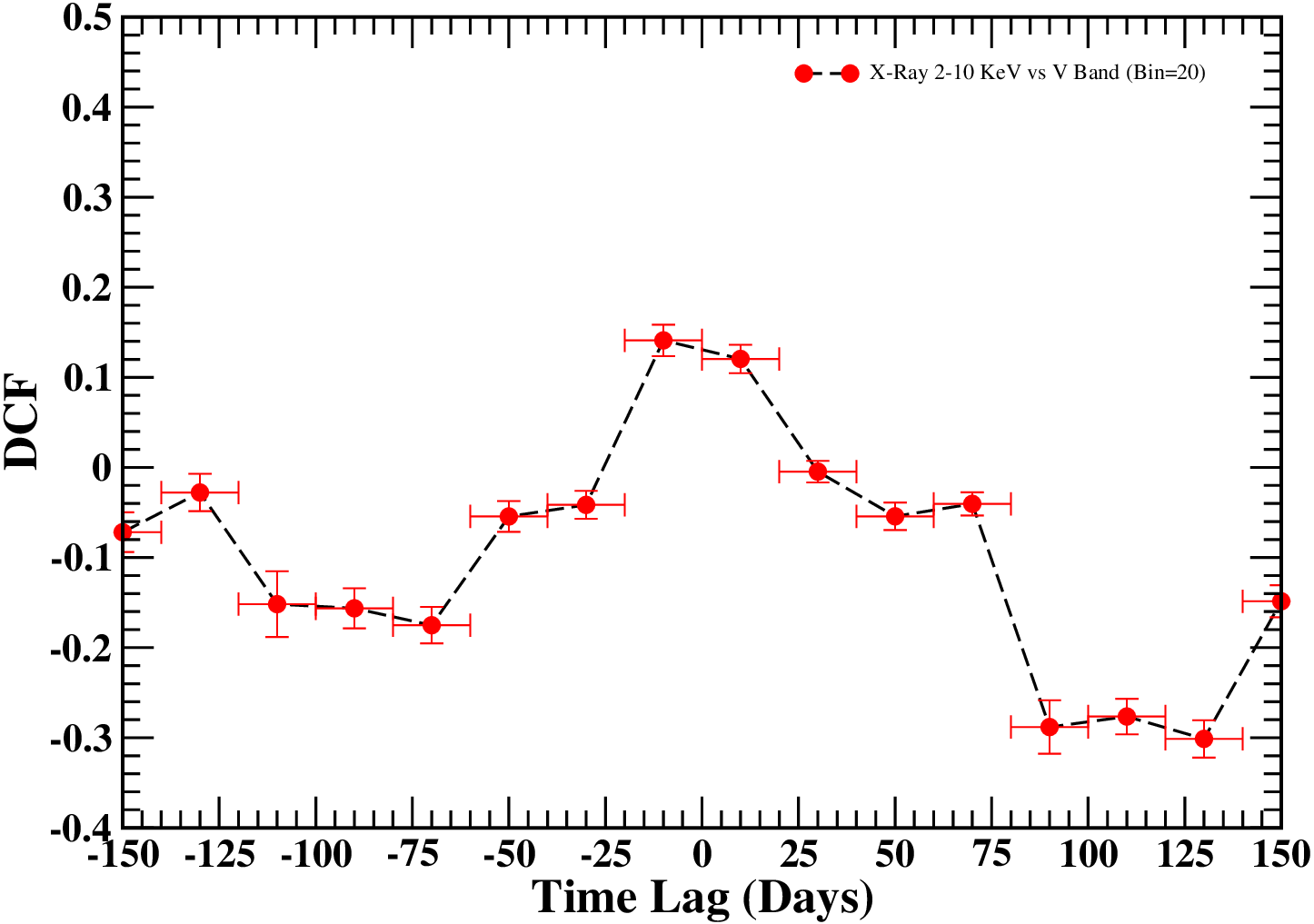}
\caption{Cross-correlation analysis curves. }
\label{fig_app_2}
\end{figure*}

\begin{figure*}[h]
\includegraphics[scale=0.18, trim=1 1 1 1.5, clip=true]{SF_2_6GHz}
\includegraphics[scale=0.18, trim=1 1 1 1.5, clip=true]{SF_5GHz}
\includegraphics[scale=0.18, trim=1 1 1 1.5, clip=true]{SF_8GHz}
\includegraphics[scale=0.18, trim=1 1 1 1.5, clip=true]{SF_10GHz}
\includegraphics[scale=0.18, trim=1 1 1 1.5, clip=true]{SF_15GHz}
\includegraphics[scale=0.18, trim=1 1 1 1.5, clip=true]{SF_23GHz}
\includegraphics[scale=0.18, trim=1 1 1 1.5, clip=true]{SF_32GHz}
\includegraphics[scale=0.18, trim=1 1 1 1.5, clip=true]{SF_43GHz}
\includegraphics[scale=0.18, trim=1 1 1 1.5, clip=true]{SF_86GHz}
\includegraphics[scale=0.18, trim=1 1 1 1.5, clip=true]{SF_142GHz}
\includegraphics[scale=0.18, trim=1 1 1 1.5, clip=true]{SF_350GHz}
\includegraphics[scale=0.18, trim=1 1 1 1.5, clip=true]{SF_Jflx}
\includegraphics[scale=0.18, trim=3 1 1 1.5, clip=true]{SF_Rflx}
\includegraphics[scale=0.18, trim=1 1 1 1.5, clip=true]{SF_Vflx}
\includegraphics[scale=0.18, trim=1 1 1 1.5, clip=true]{SF_SoftXray10to50KeV}
\includegraphics[scale=0.18, trim=1 1 1 1.5, clip=true]{SF_HardXray}
\includegraphics[scale=0.18, trim=-70 1 1 1.5, clip=true]{SF_gammaweekly}
\caption{Structure function curves at radio (2.6 to 350~GHz bands), IR (J band), optical (R and V bands), 
X-ray (10 -- 50~keV and 14 -- 192~keV), and $\gamma$-ray frequencies.}
\label{fig_app_3}
\end{figure*}

\begin{figure}
\includegraphics[scale=0.33]{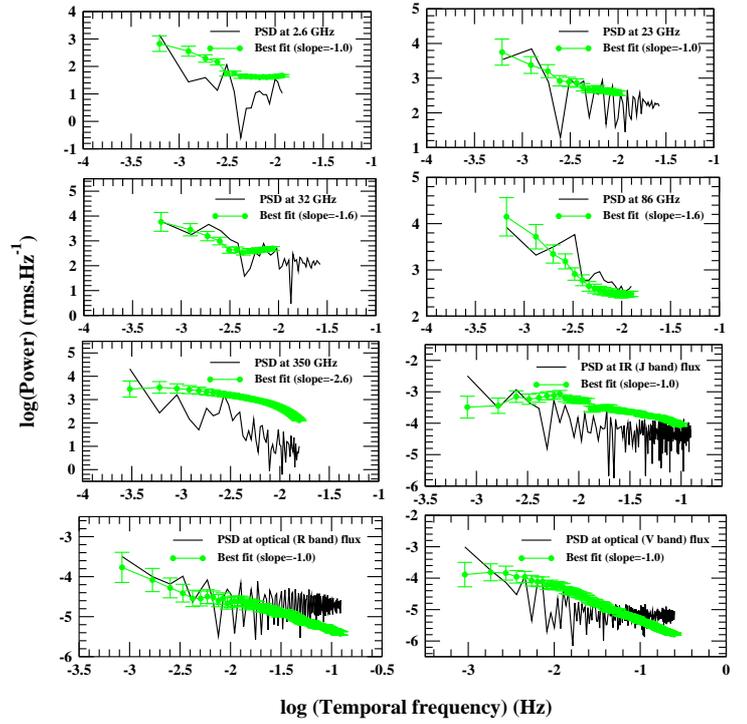}
\caption{PSD curves at radio (2.6 to 350~GHz), IR (J band), optical (R and V bands), X-ray, and $\gamma$-ray frequencies.}
\label{fig_app_4}
\end{figure}

\begin{figure}
\includegraphics[scale=0.31, trim=1 1 1 1.5, clip=true]{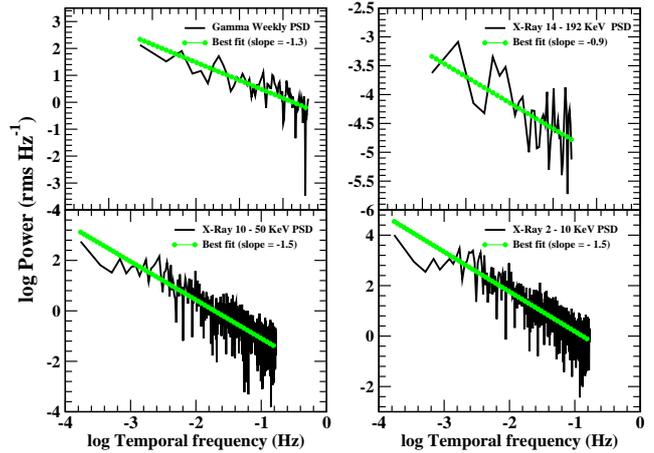}
\caption{PSD curves at X-ray and $\gamma$-ray frequencies.}
\label{fig_app_5}
\end{figure}

\end{appendix}

\end{document}